\documentclass[a4paper]{spie}  
\usepackage[]{graphicx}
\usepackage[utf8]{inputenc} 
\usepackage[T1]{fontenc} 
\usepackage{lmodern}
\usepackage[english]{babel} 
\usepackage[babel]{csquotes}
\usepackage{mathtools} 
\usepackage{amssymb} 
\usepackage{enumerate} 
\usepackage{color}

\usepackage{tikz} 
\usetikzlibrary{arrows,decorations.pathmorphing}
\usepackage{siunitx} 
\usepackage{multirow}
\usepackage{float}

\newcommand{\parderiv}[3][\empty]{\frac{\partial^{#1}#2}{\partial{#3}^{#1}}} 
\newcommand{\ima}{\operatorname{i}}
\newcommand{\e}[1]{\operatorname{e}^{\textstyle#1}}
\newcommand{\phasePl}[1]{\framebox[12pt]{$#1$}}

\definecolor{blue1}{rgb}{0,0,.7}
\definecolor{green1}{rgb}{0,.5,0}

\title{Selective higher order fiber mode excitation using a monolithic setup of a phase plate at a fiber facet} 

\author{Johannes Wilde\supit{a}, Christian Schulze\supit{a}, Robert Brüning\supit{a}, Michael Duparr\'{e}\supit{a}, Siegmund Schröter\supit{b}
\skiplinehalf
\supit{a}Institute of Applied Optics Friedrich-Schiller-University Jena, Fr\"obelstieg 1, 07743 Jena, Germany; \\
\supit{b}Leibniz Institute of Photonic Technology Jena, Albert-Einstein-Straße 9, 07745 Jena, Germany
}

\authorinfo{Further author information: \\Johannes Wilde: E-mail: johannes.wilde@uni-jena.de}

\begin{document} 
  \maketitle 

\begin{abstract}
Controlling the modal content coupled into an optical fiber can be desireable in many situations, e.g. for adjusting the sensitivity of the guided field distribution to external perturbations\cite{Sch2013}. For this purpose we used a monolithic setup of a phase plate at a fiber inpute facet to excite selectivly higher order modes, which theoretically can provide a mode purity of more than $\SI{99}{\percent}$. We investigated the capabilities of this approach by complete modal decomposition of the fiber output signals, considering the achievable mode purity with respect to several possible imperfections of the setup. The experiments are compared with detailed numerical simulations and show a high agreement.

Additionally a comparison with a well known setup with free space phase plates\cite{Sle2012,Ryf2012,Tho1994} was undertaken. This showed the monolithic setup to be energetically twice as efficient.
\end{abstract}



\section{INTRODUCTION}
Since \textit{Corning Glass Works} [known today as \textit{Corning Incorporated}] produced the first optical fiber featuring an attenuation of less than $\SI[per-mode=symbol]{20}{\decibel\per\kilo\meter}$ in $\num{1970}$, the use of optical fibers, especially in commercial communication, increased drastically. Not only in communication, but also e.g. in the fields of high power lasers for material processing\cite{Jeo2004} or minimalinvasive and robust sensors. The most common fiber sensor uses inscribed fiber Bragg gratings, that reflect certain wavelengths depending on the applied pressure, strain or current temperature\cite{Lee2002}. This setup is only sensitive at the positions, where the Bragg gratings were inscribed, and for the evaluation at least a spectral analysis is necessary if not even the demultiplexing of several signals.

In order to avoid these complicacies we wanted to utilize the quite strong dependence of higher modes guided inside a fiber on external perturbations. If only one mode is guided it should be possible to measure certain effects like strain-, pressure- or temperature-change on a long range of the fiber with a simple photo diode .

For the excitation of only one higher order mode we therefore investigated a setup of a binary phase plate directly in front of the fiber input facet [called \enquote{monolithic setup}]. From our best knowledge this monolithic setup has not yet been investigated and the following experiments serve a mere proof of principle for this setup. In order to utilize this in a real sensor additional investigations will be necessary.

\section{FUNDAMENTALS}

\subsection{LP-modes}

The fiber\footnote{\textit{Nufern LMA-GDF-25/250-M}} under investigation was a weakly guiding step index fiber. It consists of three concentric cylinders, the innermost called \enquote{core}, the next \enquote{cladding} and the outermost \enquote{coating}. The guidance of the light in the fiber can be understood by total internal reflection at the boundary surface between the core and the cladding. Therefore the refractive index difference $\Delta n = n_{\text{core}} - n_{\text{cladding}}$ has to be greater than zero. Considering the cylinders each to be non-magnetic, homogenous, linear and $\Delta n$ to be small, the wave equation for the electric field $\vec{E}$ holds\cite{Mit2005}
\begin{equation}
	\Delta \vec{E} - \frac{n^2}{c^2} \parderiv[2]{\empty}{t} \vec{E} = 0 \; ,
	\label{eq:One7}
\end{equation}

and yields for field distributions guided mostly inside the core [i.e. assuming the field to exponentially decrease in the cladding and thus neglecting effects from the coating or even more outward] with the prominent propagation direction $z$, propagation constant $\beta$ and accordingly to the problem cylindrical coordinates $(r, \varphi, z)$ for a monochromatic wave of angular frequency $\omega_0$
\begin{equation}
	\vec{E} = E_0 \, \mathcal{R}(r) \, \mathcal{P}(\varphi) \, \e{\ima [\beta z - \omega_0 t]} \, \vec{e}_{r} \; .
	\label{eq:One8}
\end{equation}

Here $E_0$ depicts the amplitude, $t$ the time and $\vec{e}_r$ an transversally arbitrarily oriented, unit polarization vector. From the smoothness at the boundary of core and cladding and the periodicity of $\varphi$ it follows [neglecting the arbitrary linear polarization]
\begin{equation}	E_{lm}^{\, p} = E_0 \begin{array}{rclcl}
	\multirow{2}{*}{\scalebox{1}[1.5]{$\Bigg\{$}} & \displaystyle \frac{1}{J_l (u_{lm})} J_l (u_{lm} \frac{r}{a})	& \multirow{2}{*}{\scalebox{1}[1.5]{$\Bigg\}$}} & \multirow{2}{*}[-8pt]{$\displaystyle \hspace*{-6pt} \cos(l \varphi + \varphi_0^{\, p}) \e{\ima [\beta_{lm} z - \omega_0 t]}$} \hspace*{.5cm} & , 0 \leq r < a	\\[9pt]
	& \displaystyle \frac{1}{K_l (w_{lm})} K_l (w_{lm} \frac{r}{a})	& & 	& , a \leq r \; 
	\end{array} \; .
	\label{eq:LPmodeField}
\end{equation}
with the core radius $a$, $u_{lm}$ and $w_{lm}$ fiber-, mode- and wavelength-dependent constants, $J_l$ the Bessel function of first kind and order $l$ and $K_l$ the modified Bessel function of second kind and order $l$ are used. These are the $LP$-modes. The index $p$ distinguishes betwenn the so called \enquote{even} and \enquote{odd} orientation of the modes.

\begin{figure}[H]						
	\centering

\scalebox{.8}{\begin{tikzpicture}[scale = 1]
	\draw (0,0) node {\includegraphics[width=.2\textwidth, trim = 4cm 6cm 5cm 8cm, clip=true]{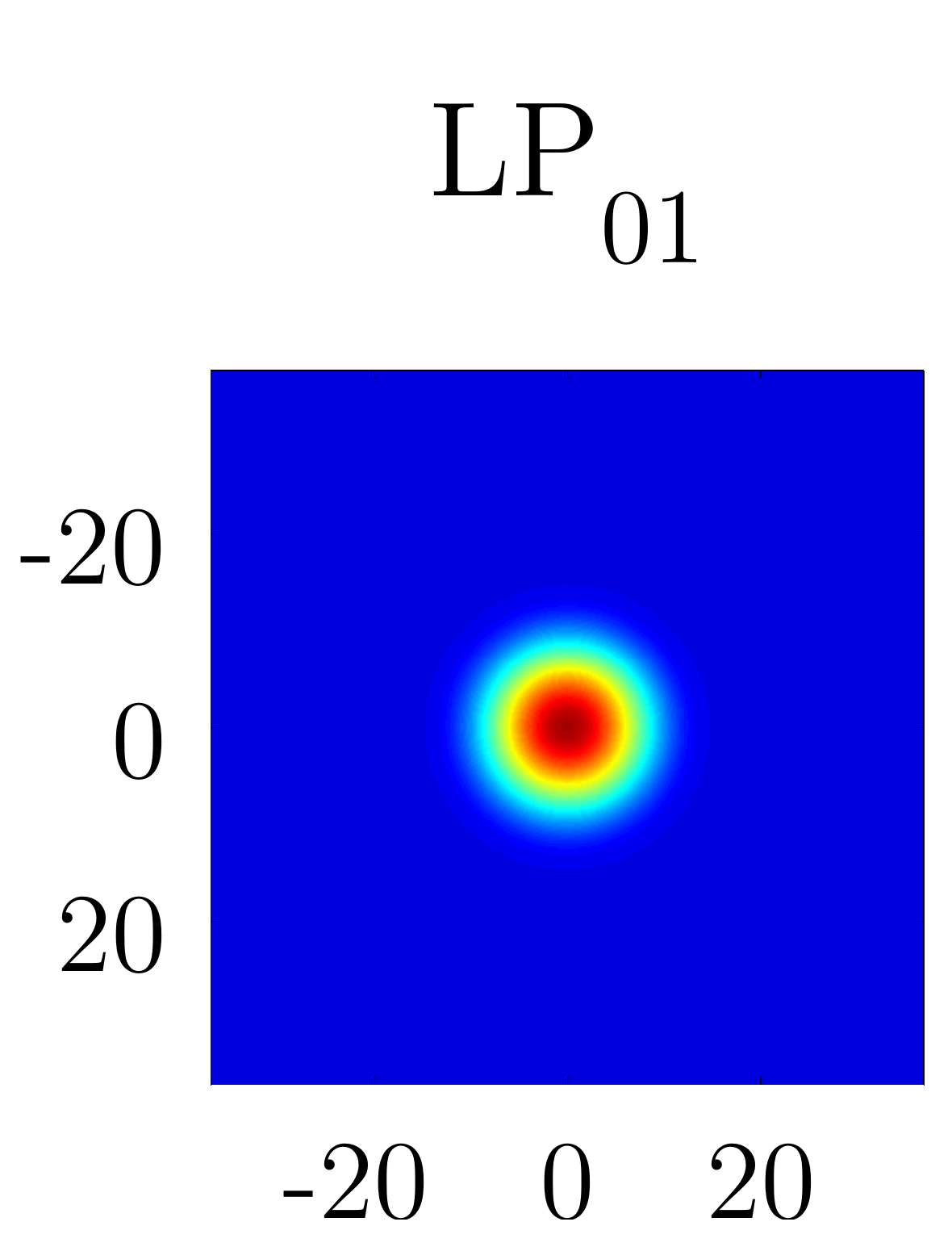}};
	\draw (1.4,0.7) node {\includegraphics[height=.06\textwidth, trim = 4.88cm 7.9cm 3.5cm 9.2cm, clip=true]{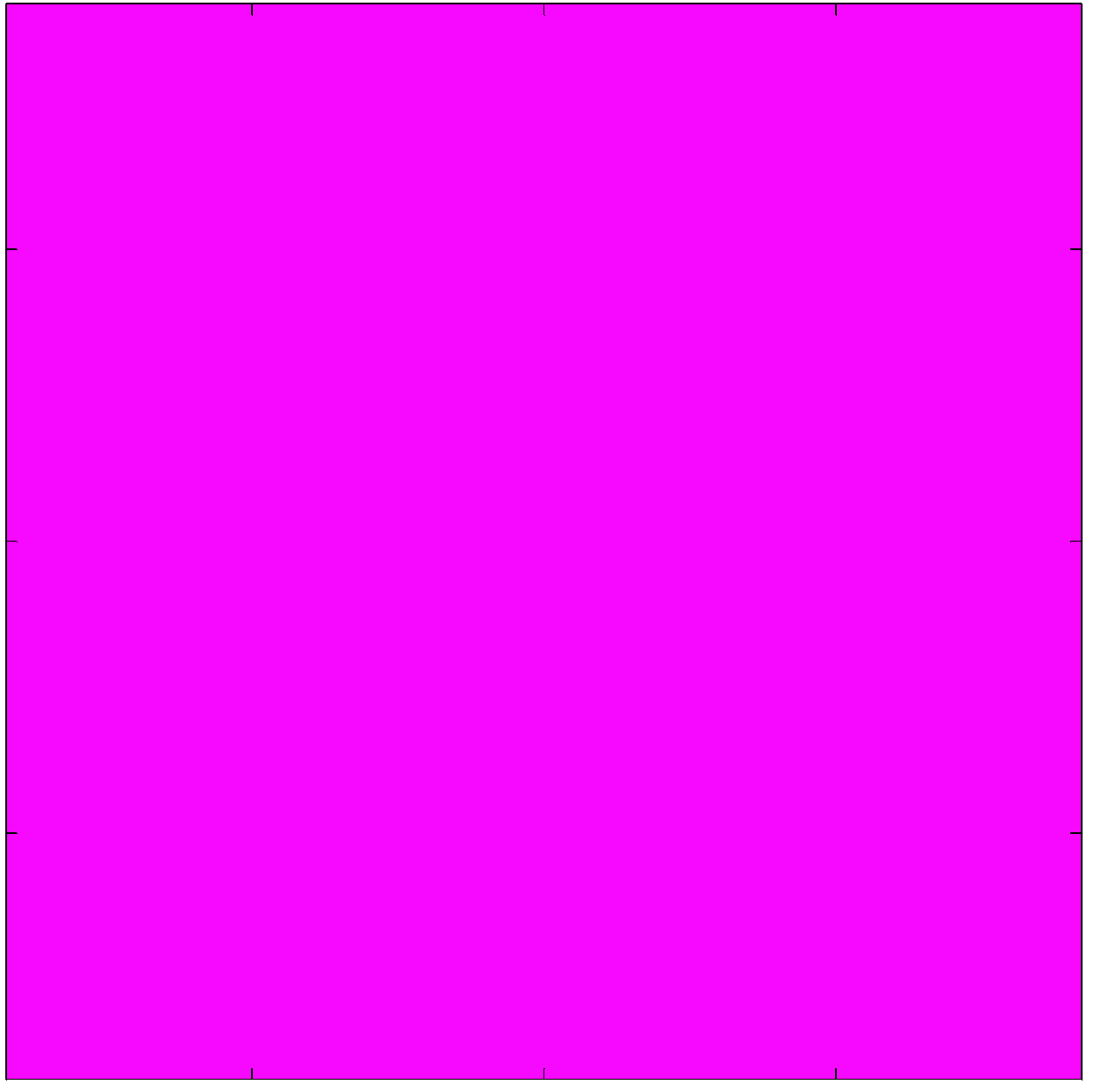}};
	
	\draw (4,0) node {\includegraphics[width=.2\textwidth, trim = 4cm 6cm 5cm 8cm, clip=true]{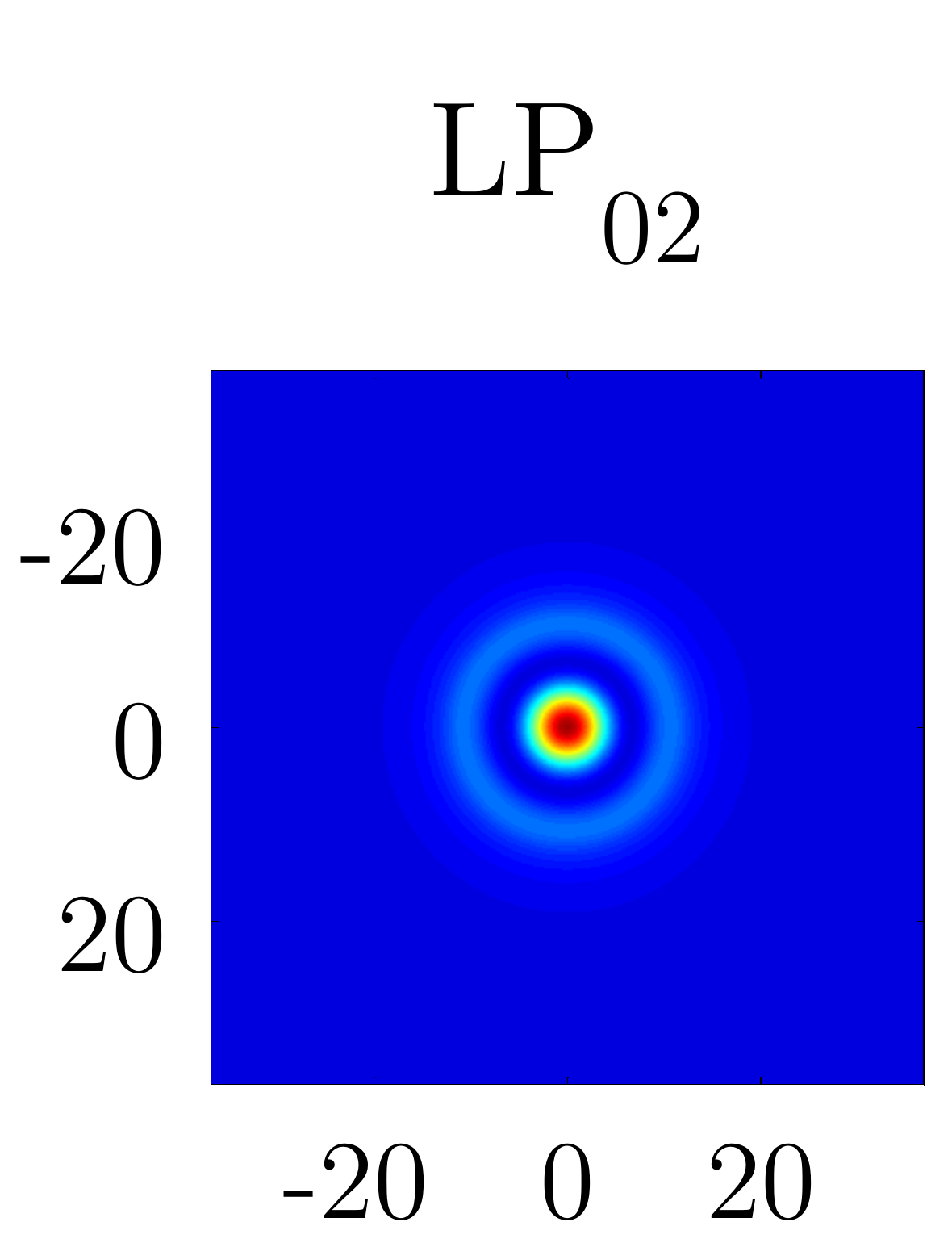}};
	\draw (4,0) ++(1.4,0.7) node {\includegraphics[height=.06\textwidth, trim = 4.88cm 7.9cm 3.5cm 9.2cm, clip=true]{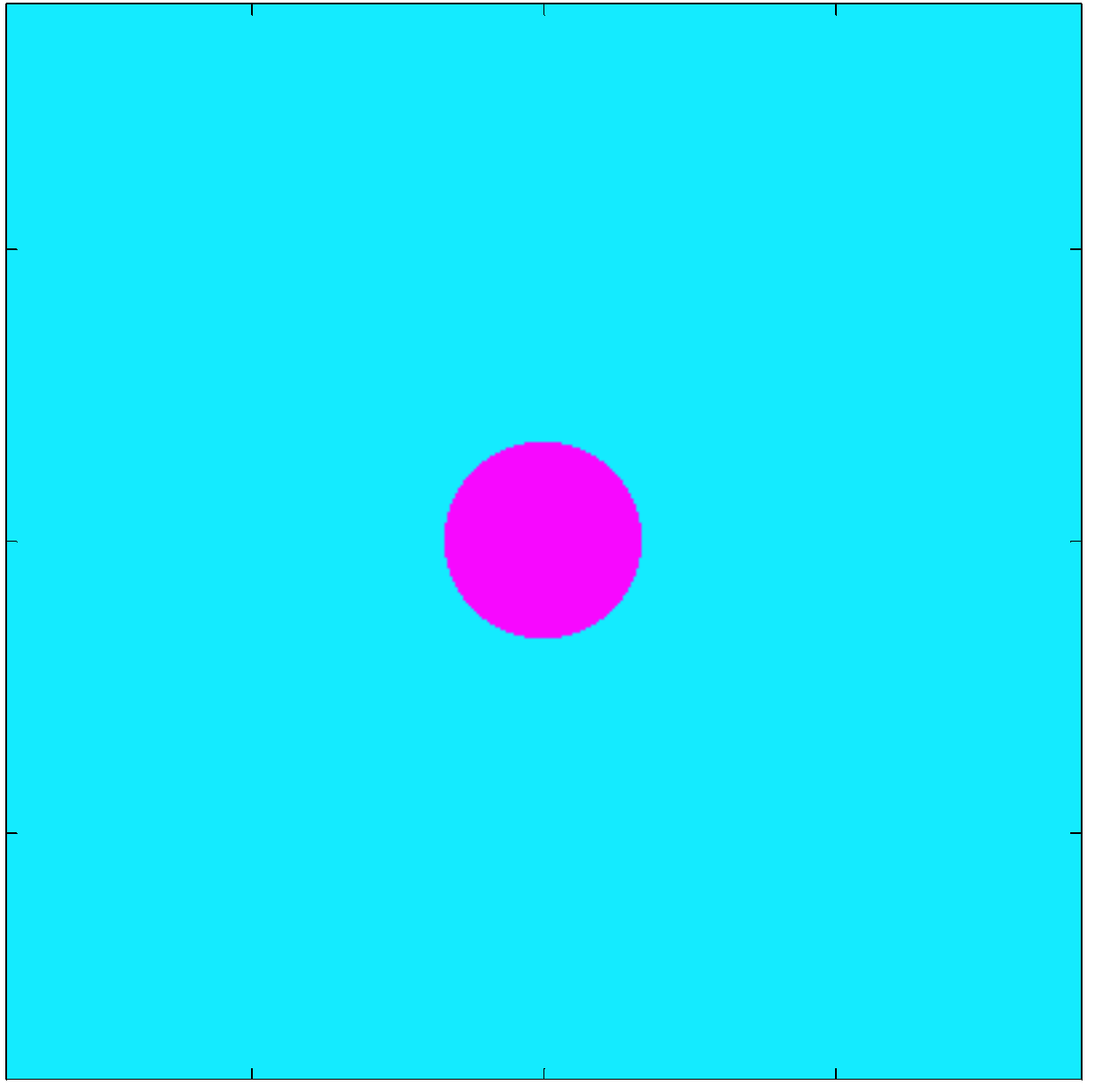}};
	
	\draw (8,0) node {\includegraphics[width=.2\textwidth, trim = 4cm 6cm 5cm 8cm, clip=true]{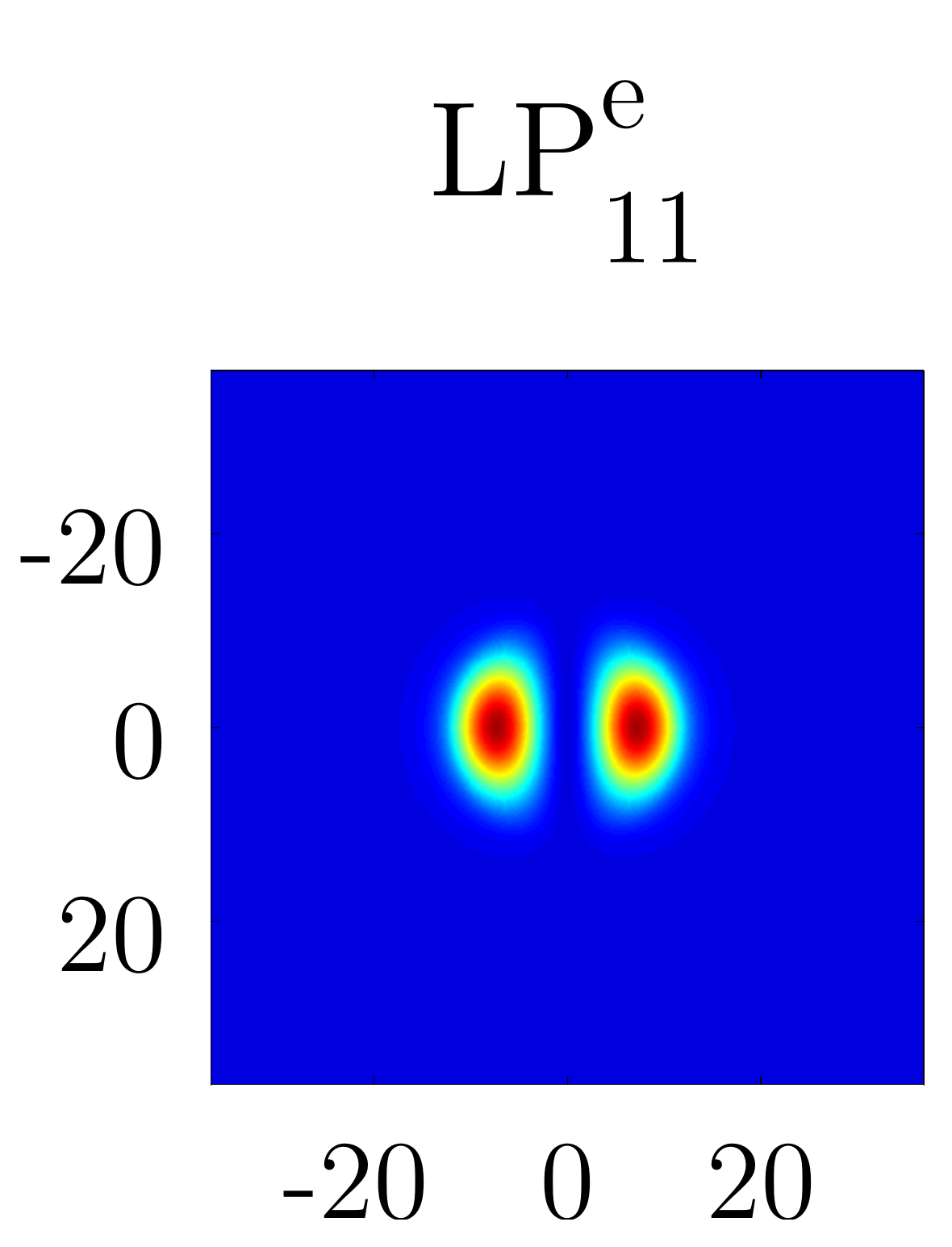}};
	\draw (8,0) ++(1.4,0.7) node {\includegraphics[height=.06\textwidth, trim = 4.88cm 7.9cm 3.5cm 9.2cm, clip=true]{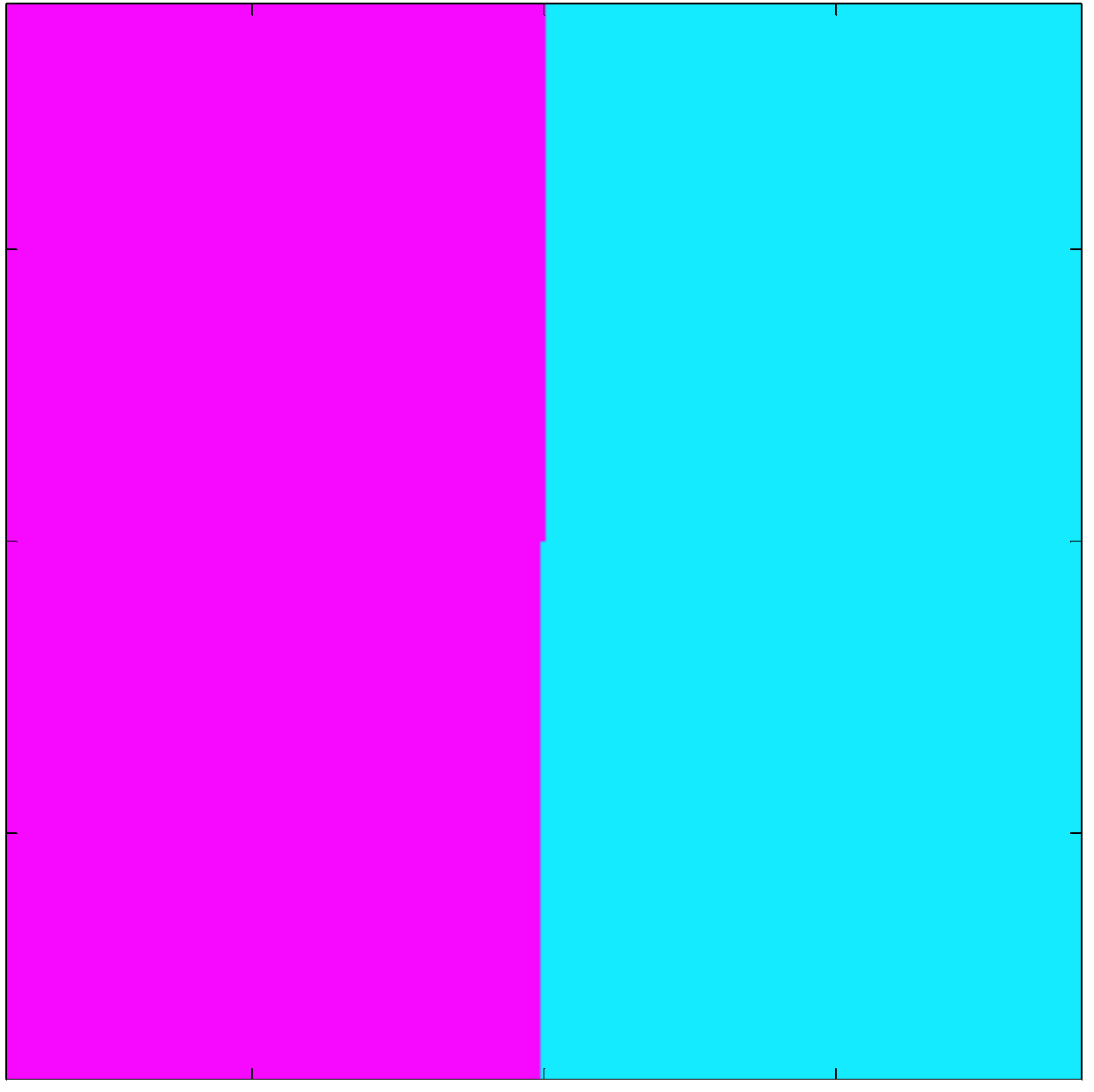}};

	\draw (11,-.2) node {\includegraphics[height=.26\textwidth, trim = 16cm 6cm 2.5cm 4cm, clip=true]{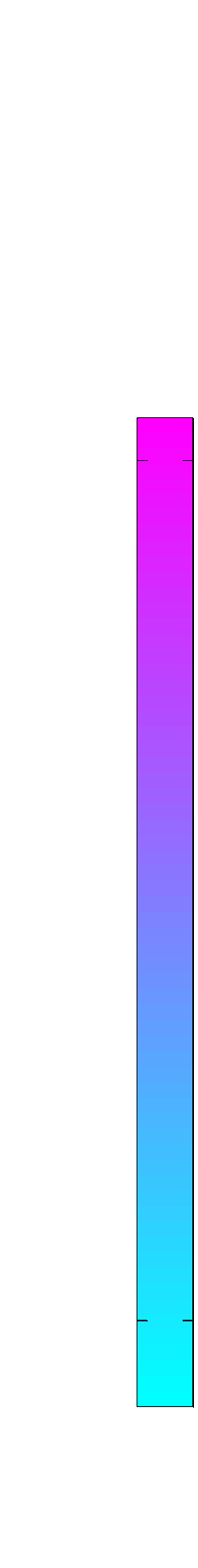}};
	\draw (11,-.2) ++(.55,-1.5) node {$\num{0}$};
	\draw (11,-.2) ++(.55,.85) node {$\pi$};

	\draw (0,0) ++(0,-5) node {\includegraphics[width=.2\textwidth, trim = 4cm 6cm 5cm 8cm, clip=true]{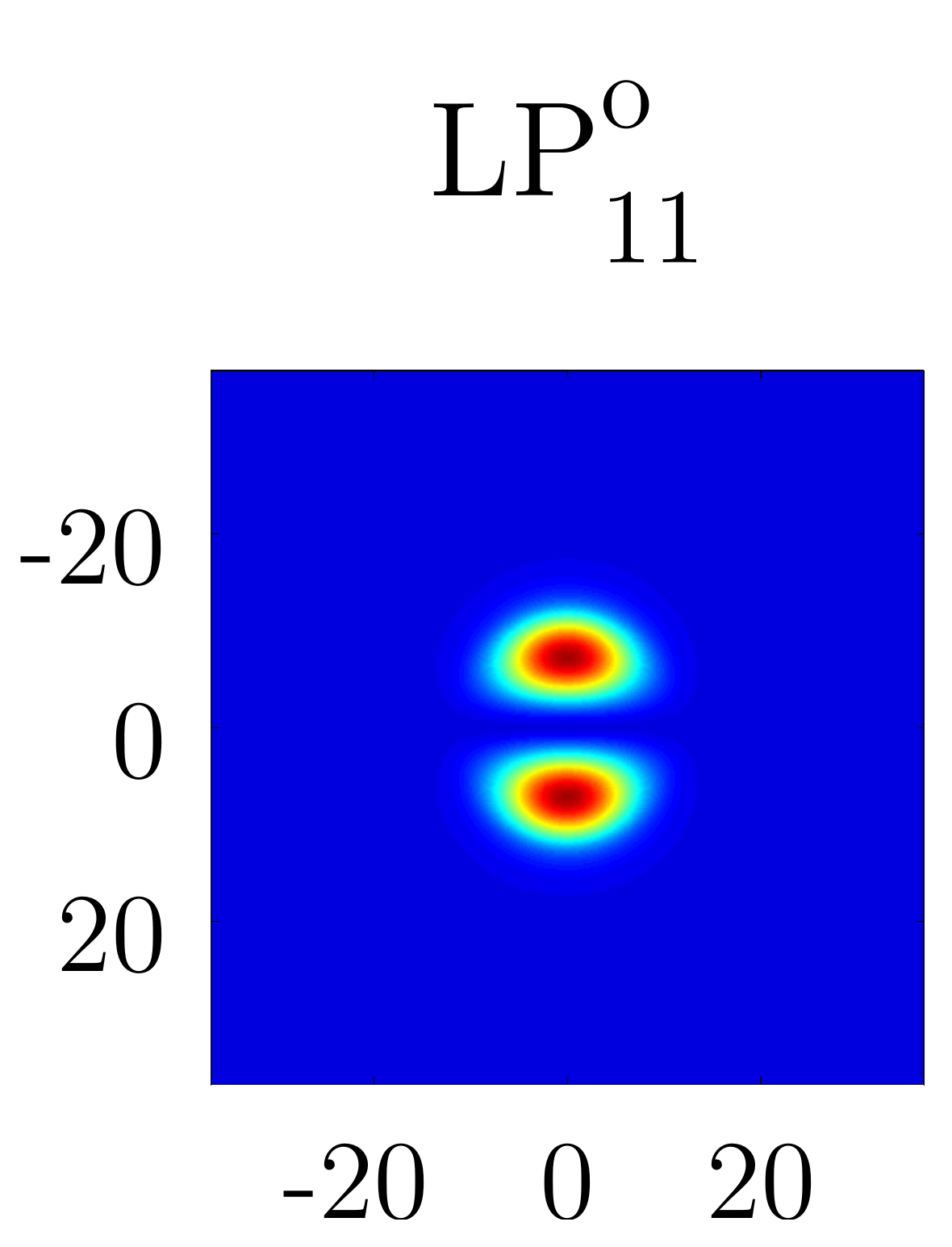}};
	\draw (1.4,0.7) ++(0,-5) node {\includegraphics[height=.06\textwidth, trim = 4.88cm 7.9cm 3.5cm 9.2cm, clip=true]{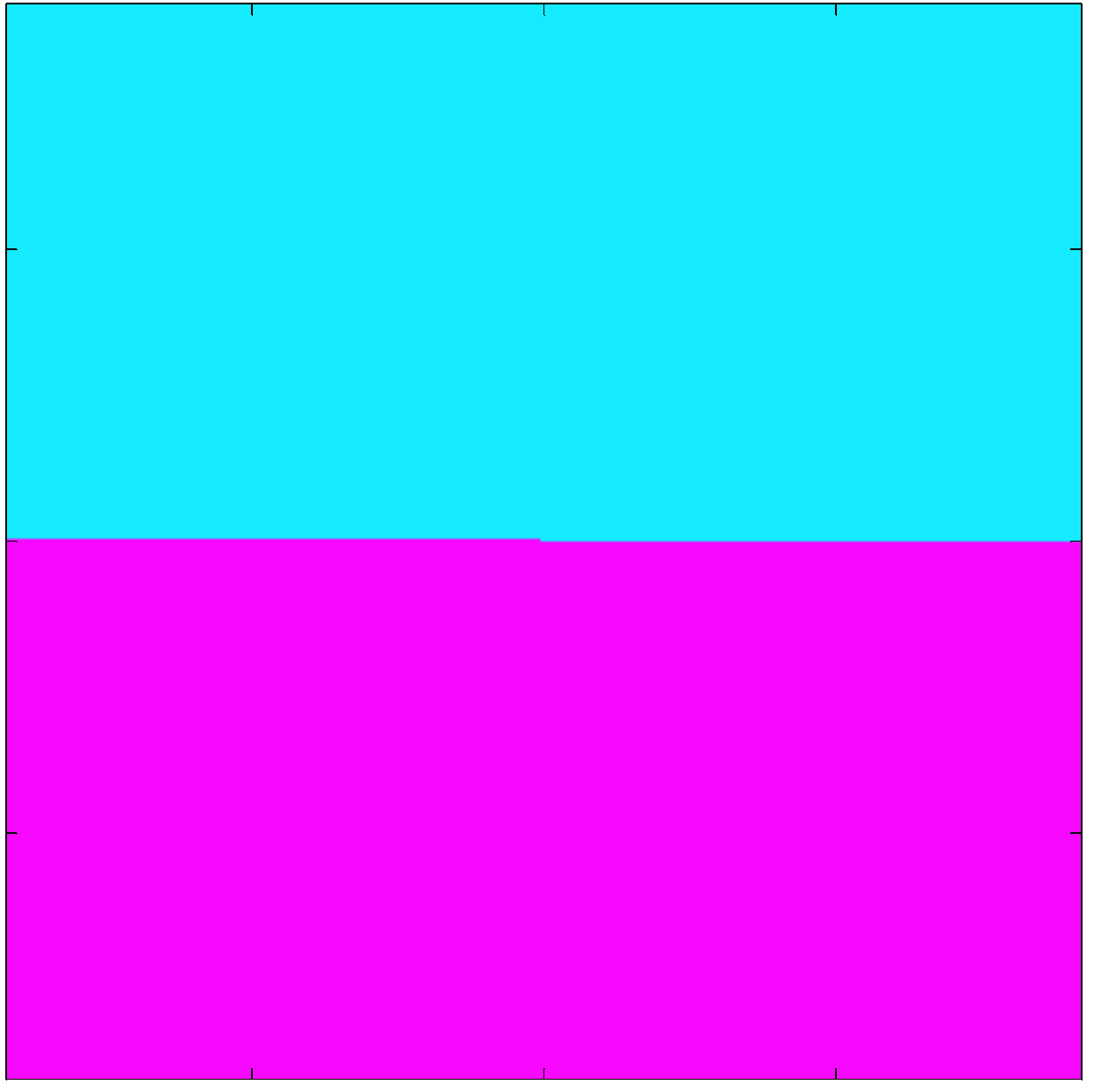}};
	
	\draw (4,0) ++(0,-5) node {\includegraphics[width=.2\textwidth, trim = 4cm 6cm 5cm 8cm, clip=true]{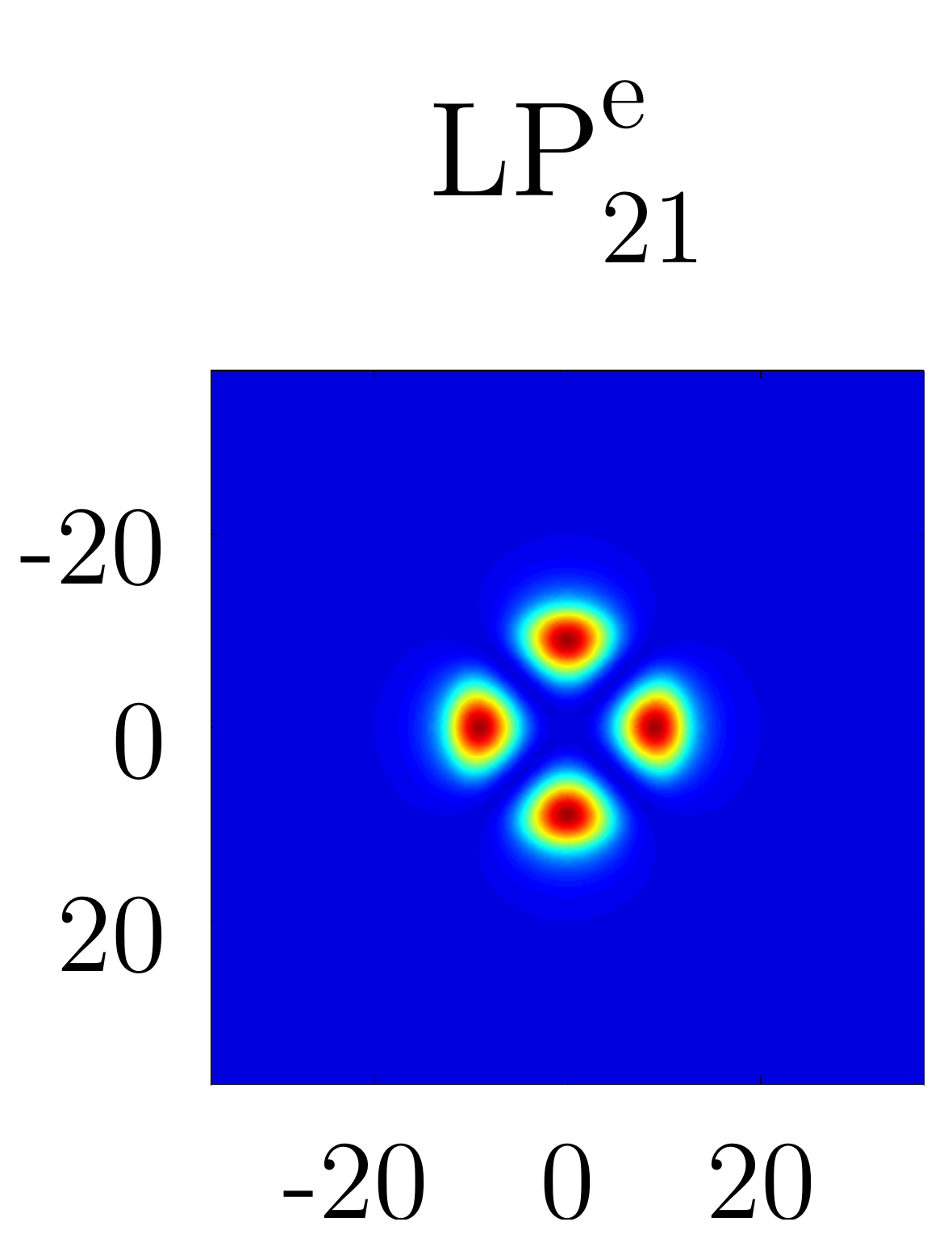}};
	\draw (4,0) ++(0,-5) ++(1.4,0.7) node {\includegraphics[height=.06\textwidth, trim = 4.88cm 7.9cm 3.5cm 9.2cm, clip=true]{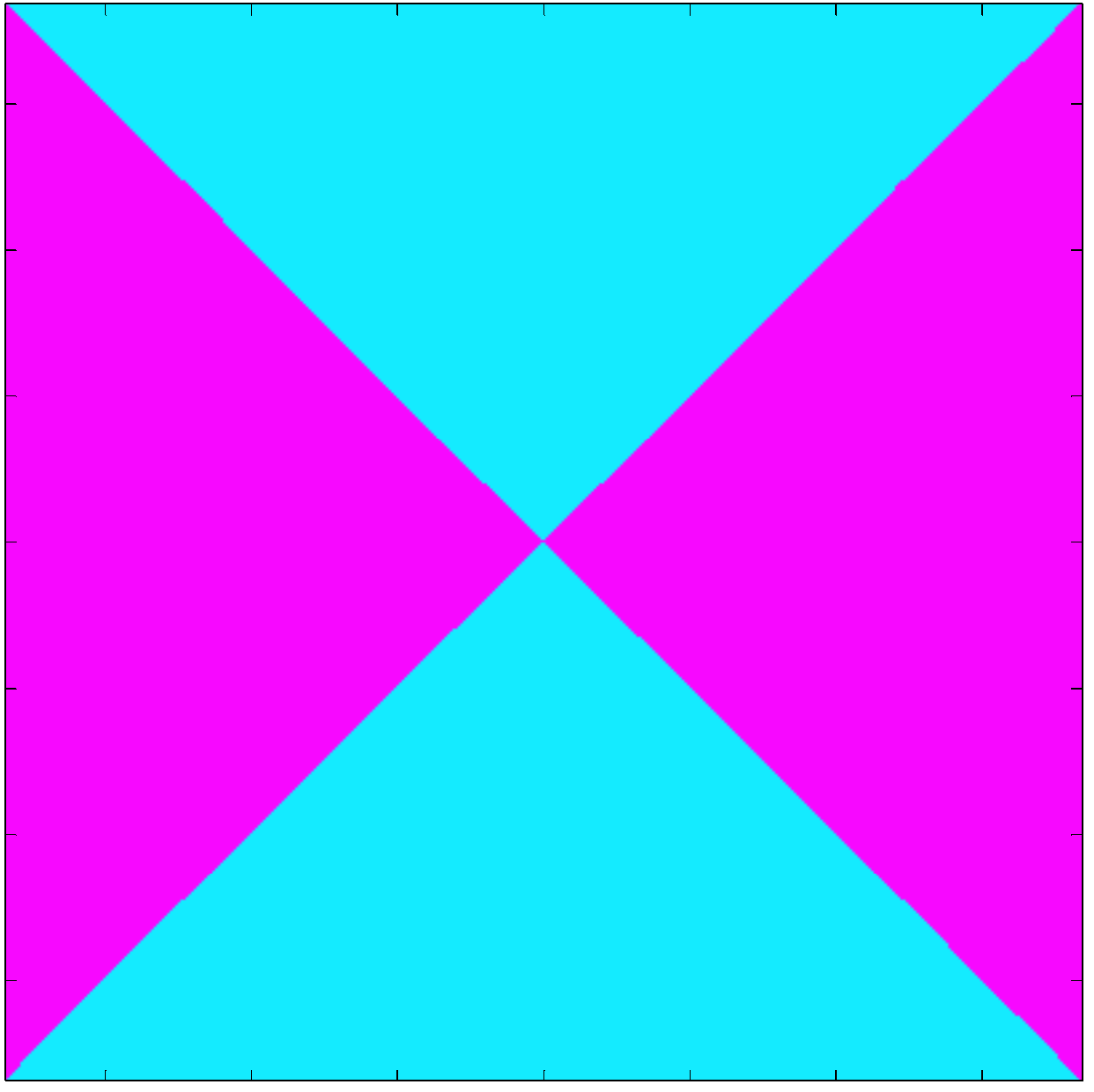}};
	
	\draw (8,0) ++(0,-5) node {\includegraphics[width=.2\textwidth, trim = 4cm 6cm 5cm 8cm, clip=true]{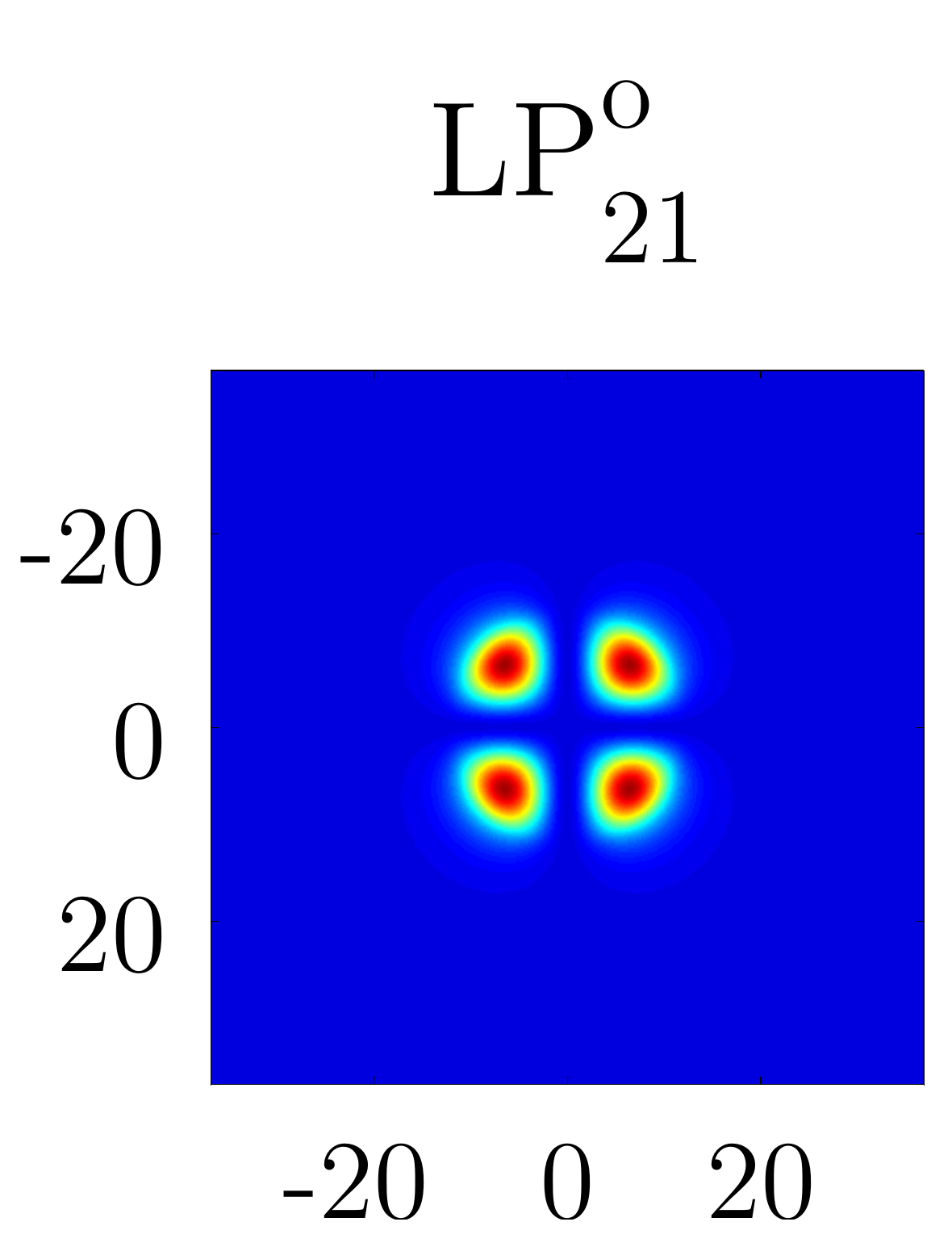}};
	\draw (8,0) ++(0,-5) ++(1.4,0.7) node {\includegraphics[height=.06\textwidth, trim = 4.88cm 7.9cm 3.5cm 9.2cm, clip=true]{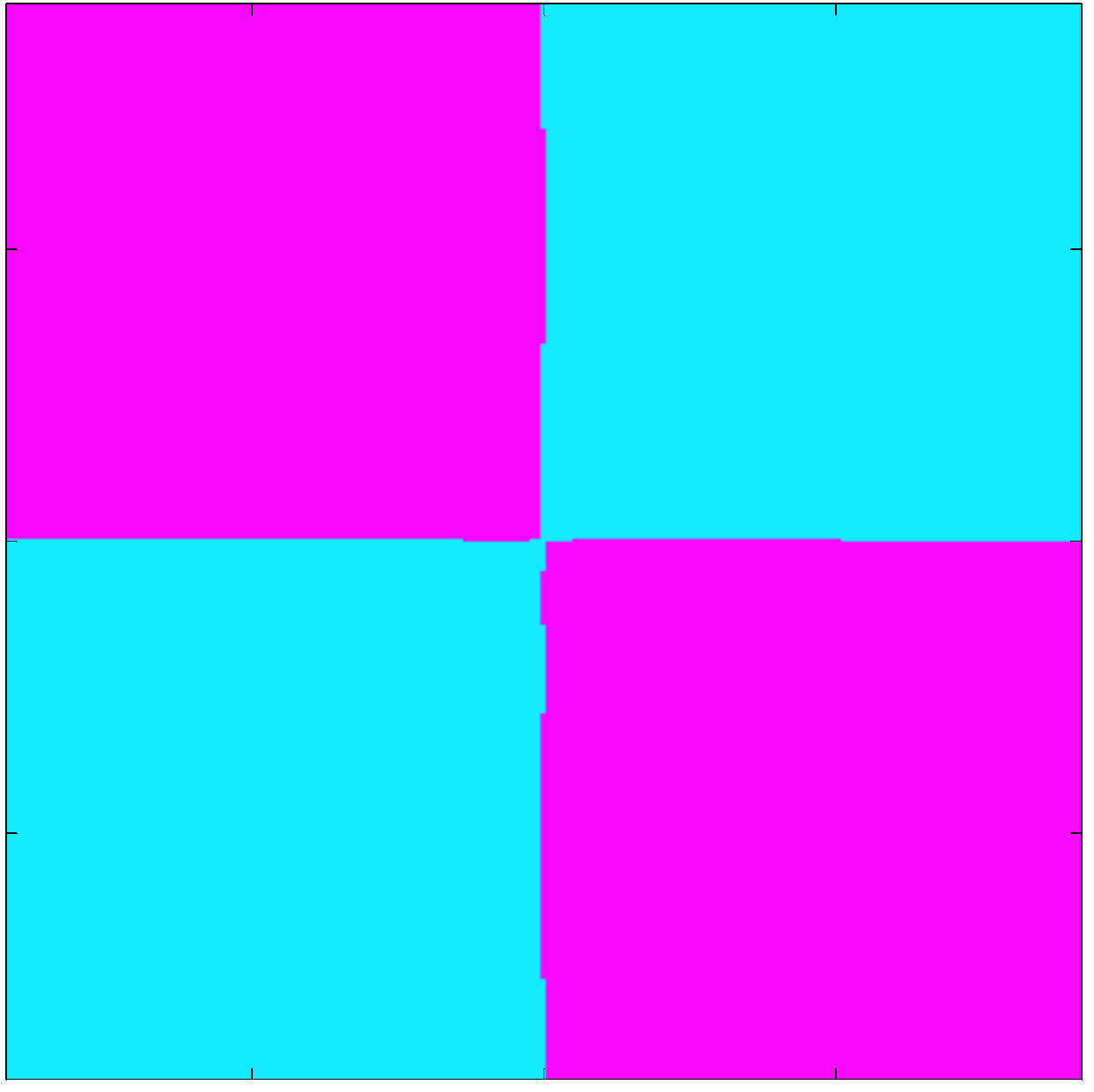}};

	\draw (11,-.2) ++(0,-5) node {\includegraphics[height=.26\textwidth, trim = 16cm 6cm 2.5cm 4cm, clip=true]{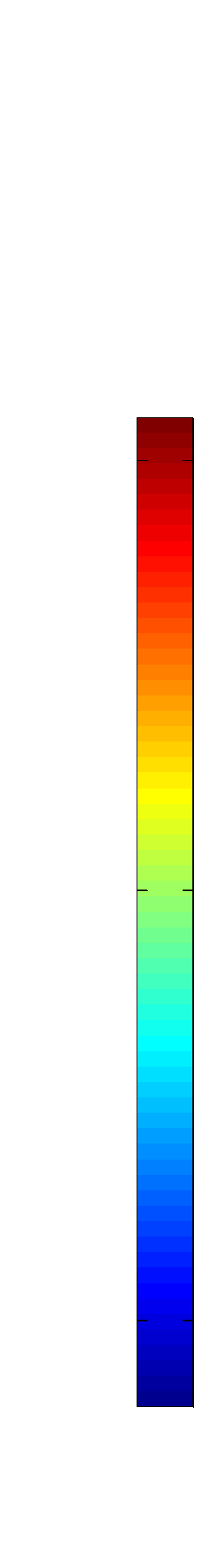}};
	\draw (11,-.2) ++(.55,-1.5) ++(0,-5) node {$\num{0}$};
	\draw (11,-.2) ++(.75,-.325) ++(0,-5) node {$\num{.5}$};
	\draw (11,-.2) ++(.55,.85) ++(0,-5) node {$\num{1}$};
	\end{tikzpicture}}

	\caption{Amplitude and phase [insets] profile of the six guided $LP$-modes in the fiber of wavelength $\lambda = \SI{1064}{\nano\meter}$ calculated numerically; dimensions in $\si{\micro\meter}$.}
	\label{fig:fibermodes}
\end{figure}
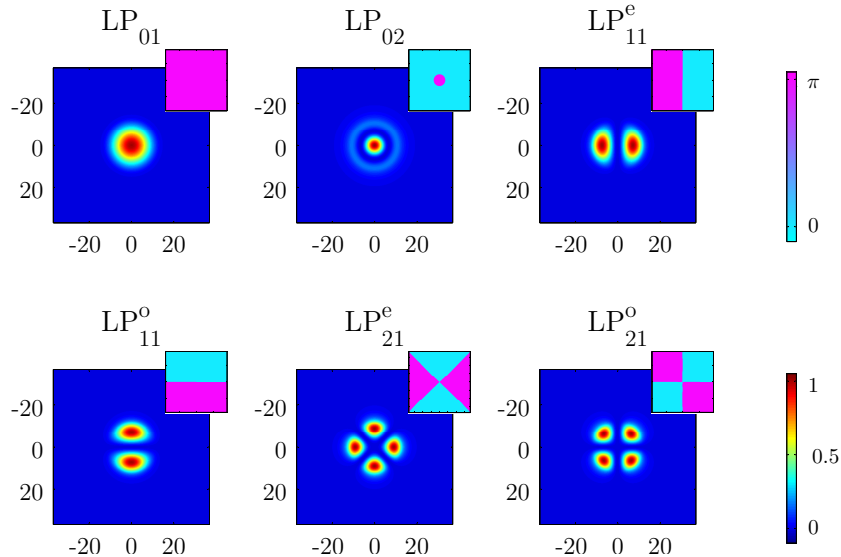

The modes calculated for the used fiber are shown in figure \ref{fig:fibermodes}.

\subsection{Binary Passive Phase Plates}

\begin{figure}[H]
\centering
\begin{tikzpicture}[scale=.22]
	\filldraw[fill=red!15, draw=red!10] (1,2.25) -- (11,2.25) -- (11,-.5) -- (1,1.5);	
	\filldraw[fill=red!25, draw=red!10] (1,3) -- (11,5) -- (11,2.25) -- (1,2.25);	
	\filldraw[fill=white, draw=white] (1,3) .. controls (5,3.5) and (8,4).. (11,5);
	\filldraw[fill=white, draw=white] (1,1.5) .. controls (5,1) and (8,.5).. (11,-.5);
	\draw[draw=white, thick] (1,3) -- (11,5) (11,-.5) -- (1,1.5);
	\draw[draw=black!70] (1,3) .. controls (5,3.5) and (8,4).. (11,5);
	\draw[draw=black!70] (1,1.5) .. controls (5,1) and (8,.5).. (11,-.5);
	\draw[color=black!80] (5,2.25) node {$E_{\text{in}}$};
	
	\draw[color=white, very thick] (2,-1) -- ++(0,2.35) -- ++(-.5,0) -- ++(0,.9) -- ++(.5,0) -- ++(0,3.25);
	
	\filldraw[fill = black!20, draw=black!20] (0,-1) -- (2,-1) -- (2,1.35) -- (1.5,1.35) -- (1.5,2.25) -- (2,2.25) -- (2,5.5) -- (0,5.5) -- (0,-1); 
	\draw[thick] (2,-1) -- (2,1.35) -- (1.5,1.35) -- (1.5,2.25) -- (2,2.25) -- (2,5.5);
	\draw[thick] (0,-1) -- (0,5.5);
	
	\draw[thick] (1,4) -- (-1,6) node[anchor = south east] {$n_{\text{ph}}$};
	\draw[->, thick] (.5,6) -- (1.5,6);
	\draw[->, thick] (3,6) -- (2,6);
	\draw (1.75,6) node[above] {$d$};
	
	\filldraw[fill=red!25, draw=red!10] (-.1,3) -- (-10,5) -- (-10,-.5) -- (-.1,1.5);	
	\filldraw[fill=white, draw=white] (-.1,3) .. controls (-4,3.5) and (-7,4).. (-10,5);
	\filldraw[fill=white, draw=white] (-.1,1.5) .. controls (-4,1) and (-7,.5).. (-10,-.5);
	\draw[draw=white, thick] (-.1,3) -- (-10,5) (-10,-.5) -- (-.1,1.5);
	
	\draw[draw=black!70] (-.1,3) .. controls (-4,3.5) and (-7,4).. (-10,5);
	\draw[draw=black!70] (-.1,1.5) .. controls (-4,1) and (-7,.5).. (-10,-.5);
	\draw[color=black!80] (-5,2.25) node {$E_{\text{0}} \; \; \longrightarrow$};
	
	\draw[thick] (2,-3) -- (2,-2) (2,-2.5) -- (7,-2.5) (7,-3) -- (7,-2) (4.5,-2.5) node[anchor=south] {$\zeta$};
	
	\draw (12.5,5.5) node[anchor=south] {fiber};
	\filldraw[fill=black!10, draw=black!10] (7,3.25) rectangle (18,5.5); 
	\filldraw[fill=black!10, draw=black!10] (7,1.25) rectangle (18,-1); 
	\filldraw[fill=red!25, draw=red!25] (7,2.25) rectangle (18,3.25);		
	\filldraw[fill=red!15, draw=red!15] (7,1.25) rectangle (18,2.25);		
	
	\draw[draw=black!70] (18,5.5) -- (7,5.5) -- (7,3.25) -- (18,3.25) (7,3.25) -- (7,1.25) -- (18,1.25) (7,1.25) -- (7,-1) -- (18,-1);
	
	\draw[color=black!80] (13,2.25) node {$E_{\text{ein}} \; \; \longrightarrow$};
\end{tikzpicture}
\caption{Principle setup of the monolithic binary phase plate in front of the fiber input facet.}
\label{fig:BinaryPhasePlateSideView}
\end{figure}
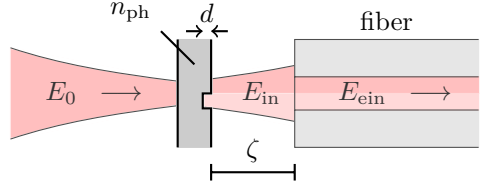

As it can be seen in figure \ref{fig:fibermodes}, the respective phase profiles of two arbitrary modes are orthogonal to each other. Therefore the phase profile of an incoming fundamental Gaussian beam was phase shaped according to these. This was accomplished by placing a passive element of refractive index $n_{\text{ph}}$ and transversally varying thickness into the focused beam waist very near to the fiber input facet. This is shown in figure \ref{fig:BinaryPhasePlateSideView}.

The free space propagation distance $\zeta$ between the phase plate and the fiber input is depicted in figure \ref{fig:BinaryPhasePlateSideView} as well.

As the refractive index of the phase plate deviates from the index of the environment $n$, a laterally varying phase shift

\begin{equation}
	\Delta \Phi = [n_{ph} - n] \frac{d}{\lambda_0} 2 \pi
	\label{eq:DeltaPhi}
\end{equation}

is induced.

\begin{figure}[H]
\centering
\begin{tikzpicture}[scale=1]
	\draw (0,0) node {\includegraphics[height=2.5cm,trim=4cm 13cm 3cm 13cm, clip]{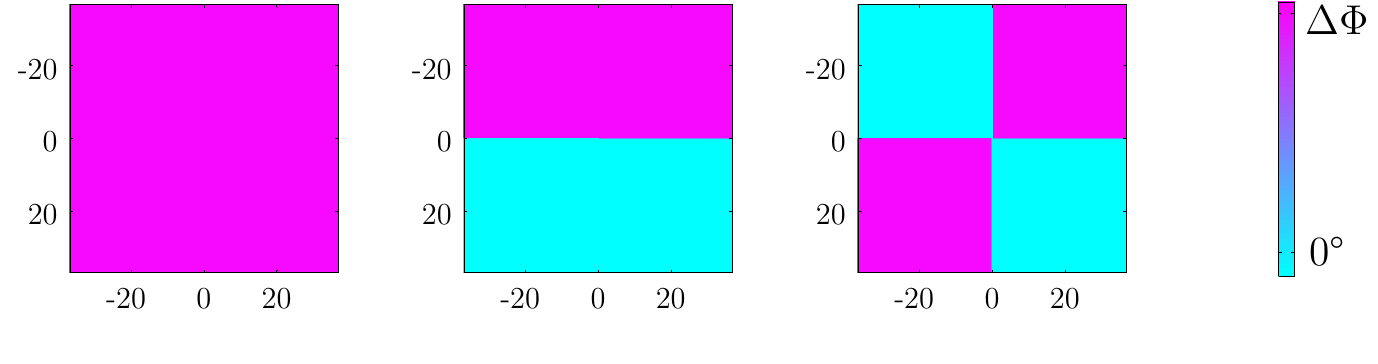}};
	\draw (-3.3,1.7) node {\small \phasePl{0}};
	\draw (-.6,1.7) node {\small \phasePl{1}};
	\draw (2.1,1.7) node {\small \phasePl{2}};
\end{tikzpicture}
\caption{Transversal profiles of the used phase plates; dimensions in $\si{\micro\meter}$.}
\label{fig:BinaryPhasePlateTransversalView}
\end{figure}

Corresponding to the modal phase profiles depicted in figure \ref{fig:fibermodes}, the phase profiles shown in figure \ref{fig:BinaryPhasePlateTransversalView} were realized and investigated.

It should be noted that the maximal phase shift depicted in figure \ref{fig:BinaryPhasePlateTransversalView} is not $\SI{180}{\degree}$. This originated from the not exactly known refractive index of the used material\footnote{Polydimethylsiloxane} for the phase plates.

It could be shown analytically that, if a perfect fundamental Gaussian beam was irradiated, the intensity percentage $\tilde{\rho}_l^{\: 2}$ of the field after the phase plate \, perfectly adapted to the modal field distribution of the respective desired mode $l$ inside the fiber relative to the total incoupled intensity, varied in the following way:

\begin{equation}
	\tilde{\rho}_l^{\: 2} = \frac{1}{2} \, [1 - \cos(\Delta \Phi)] \hspace*{1cm} , \hspace*{1cm} \tilde{\rho}_0^{\: 2} = \frac{1}{2} [1 + \cos(\Delta \Phi)] \; \; ;
	\label{eq:Two5}
\end{equation}

here $\tilde{\rho}_0^{\: 2}$ is the percentage of the incoupling intensity coupled into the $LP_{0m}$ mode group.

\section{EXPERIMENTS}

\subsection{Experimental setup}

\begin{figure}[H]
\centering
\includegraphics[width=.8\textwidth]{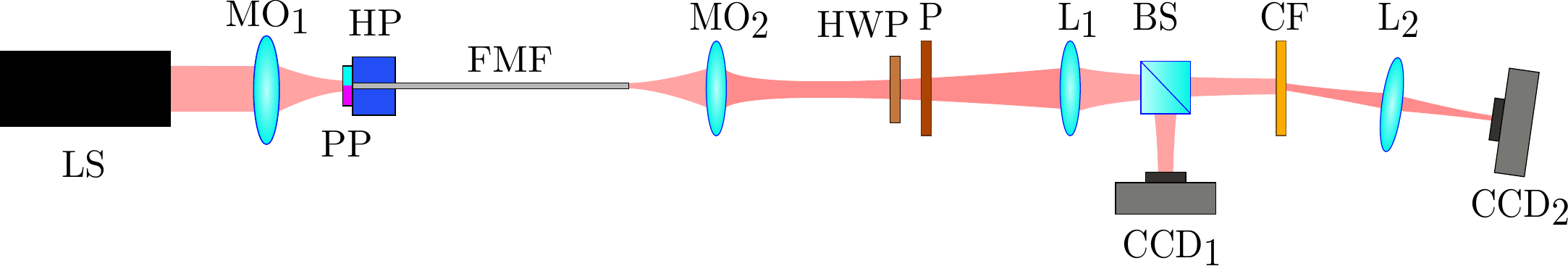}
\caption{Scheme of the experimental setup \; [ LS - laser source, \, $\text{MO}_{1,2}$ - microscope objectives, PP - phase plate, \, HP - hexapod, \, FMF - few mode fiber, \, HWP - $\frac{3 \lambda}{2}$ waveplate, P - polarizer, \, $\text{L}_{1,2}$ - lenses, \, BS - beam splitter, \, $\text{CCD}_{1,2}$ - cameras, \, CF - correlation filter].}
\label{fig:experimentalSetup}
\end{figure}

A scheme of the experimental setup is shown in figure \ref{fig:experimentalSetup}. 

On the left the collimated laser beam of wavelength $\lambda = \SI{1064}{\nano\meter}$ is shown, which is focused via a microscope objective on the phase plate and then coupled into the fiber. The fiber input facet with the phase plate was mounted on a nano positioning device [the hexapod] and thus could be aligned to the incoming beam. Additionally the fiber could be moved out of the justified position and thus the effects of a transversal misalignment investigated.

In order to investigate the modal content, coupled into the fiber, the correlation filter method [CFM]\cite{Kai2008} was employed. Together with the half wave plate and the polarizer the two transversal linear polarization states could be distinguished. Due to \enquote{modal birefringence} and \enquote{random coupling}\cite{Kog2012} of the guided fiber modes however these distinctions showed properties of the respectively used fiber and no additional information regarding the incoupling process. Therefore these experimentally made distinctions later on were neglected.

\subsection{Numeric Simulations}

The properties of the setup investigated numerically prior to the experiments were the dependence of the modal coupling on the beam waist radius $\sigma$ of the incoming beam, \, the free space propagation distance between the phase plate and the fiber input facet $\zeta$ \, and the transversal displacement of the incoming beam relative to the concentric position of incident beam and fiber.

\begin{figure}[H]
\centering
\begin{tikzpicture}[scale=1.4]
\draw (0,0) node {\includegraphics[width=.35\textwidth, trim = .1cm 4.45cm 1.5cm 8.6cm, clip=true]{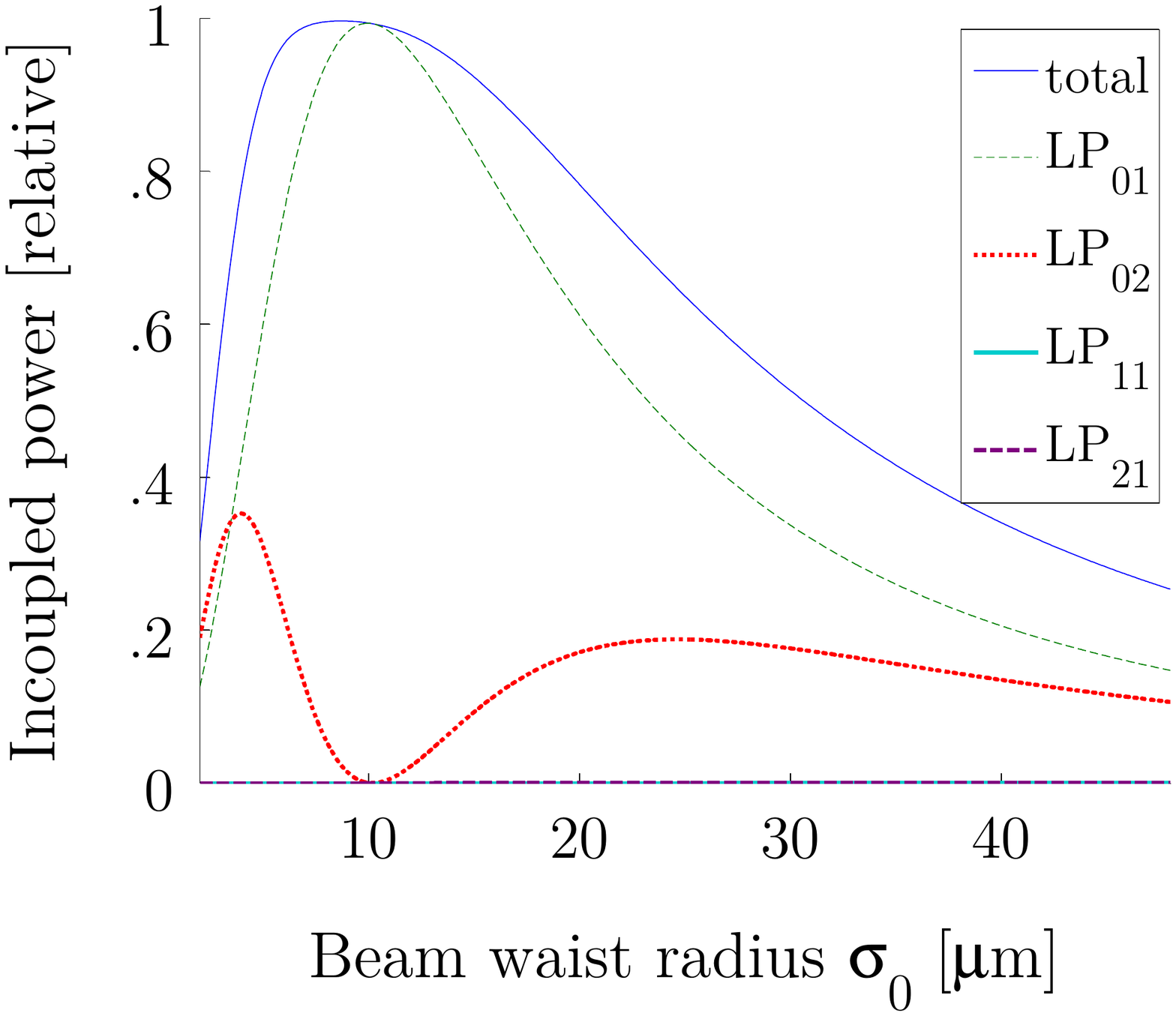}};
\filldraw[color=white] (-2,-1.5) rectangle (2,-2);
\draw (.35,-1.75) node {\footnotesize Beam waist radius $\sigma$ [$\si{\micro\meter}$]};
\filldraw[color=white] (-1.9,-1.25) rectangle (-2.5,1.7);
\draw[] (-2.1,.25) node[rotate=90] {\footnotesize Relat. incoupled intensity};
\end{tikzpicture}
\caption{Modal intensity coupled into the fiber relative to the intensity of the incident beam for phase plate \protect \phasePl{0} [$\Delta\Phi$ arbitrary].}
\label{fig:BeamSize0}
\end{figure}

\begin{figure}[H]
\centering
\hspace*{.75cm} $\Delta\Phi = \SI{180}{\degree}$ \hspace*{5.25cm} $\Delta\Phi = \SI{144}{\degree}$\\
\begin{tikzpicture}[scale=1.4]
\draw (0,0) node {\includegraphics[width=.35\textwidth, trim = .1cm 4.45cm 1.5cm 8.6cm, clip=true]{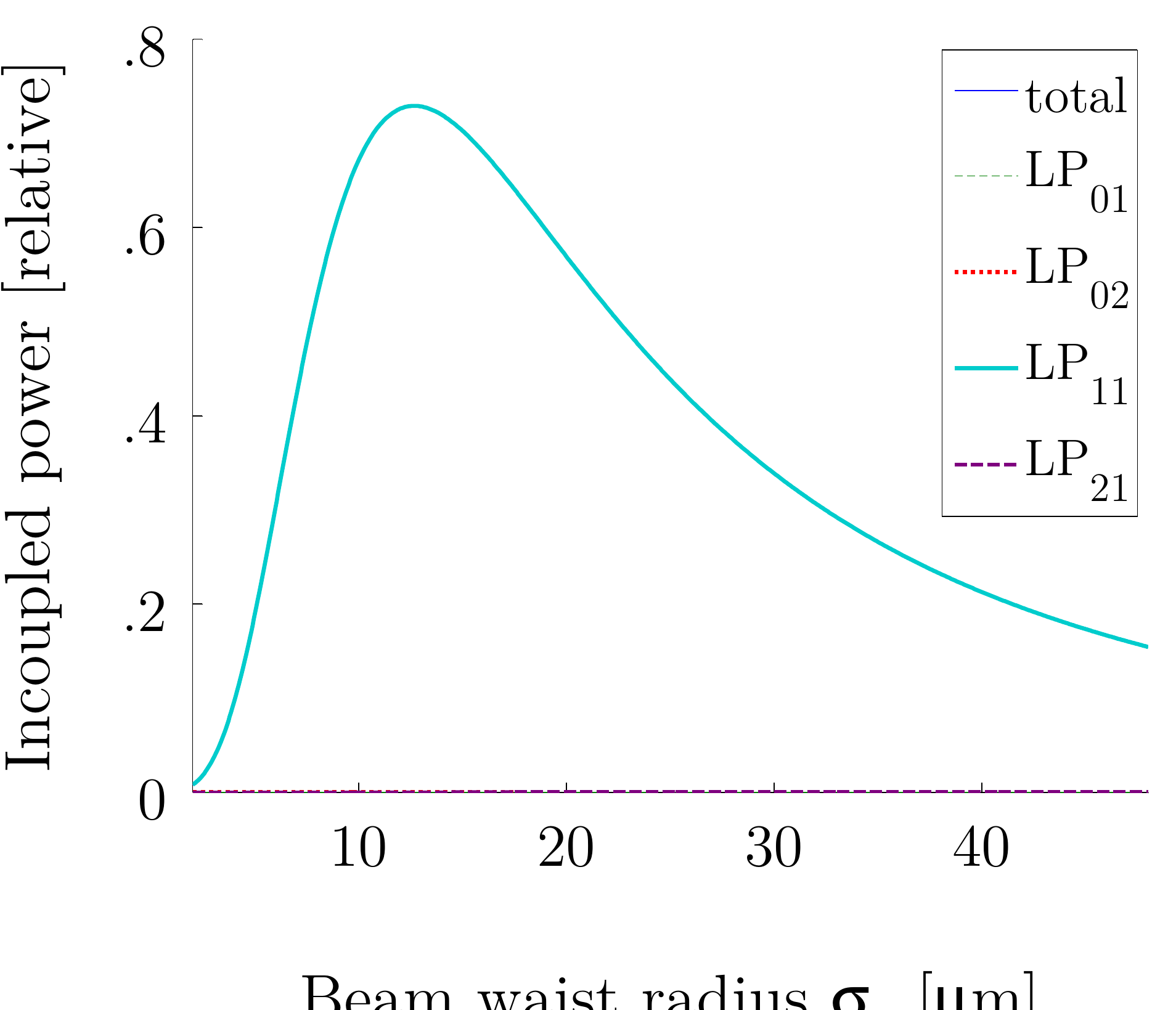}};
\filldraw[color=white] (-2,-1.5) rectangle (2,-2);
\draw (.35,-1.75) node {\footnotesize Beam waist radius $\sigma$ [$\si{\micro\meter}$]};
\filldraw[color=white] (-1.9,-1.25) rectangle (-2.5,1.7);
\draw[] (-2.1,.25) node[rotate=90] {\footnotesize Relat. incoupled intensity};
\end{tikzpicture} \hspace*{.5cm} \begin{tikzpicture}[scale=1.4]
\draw (0,0) node {\includegraphics[width=.35\textwidth, trim = .1cm 4.45cm 1.5cm 8.6cm, clip=true]{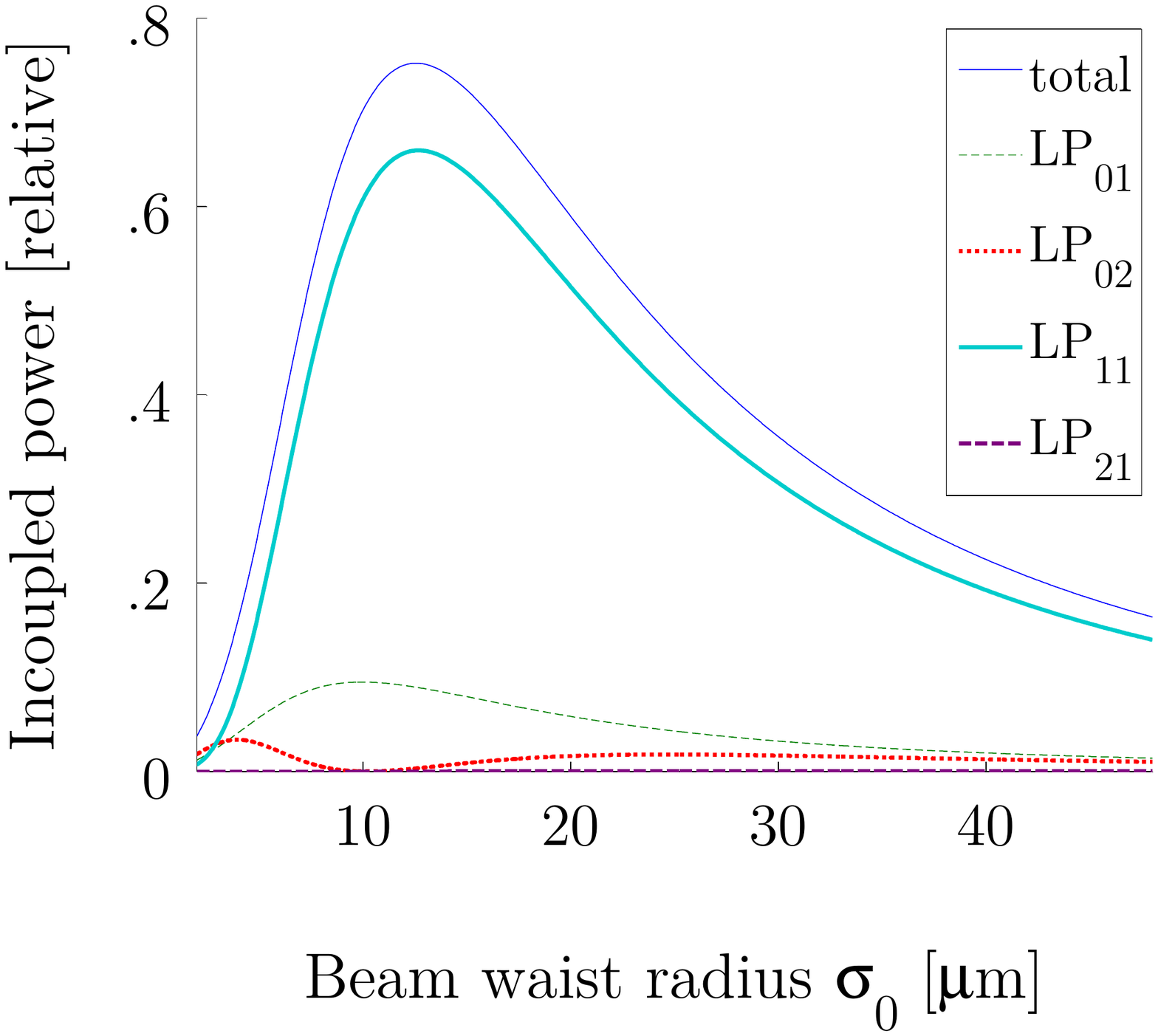}};
\filldraw[color=white] (-2,-1.5) rectangle (2,-2);
\draw (.35,-1.75) node {\footnotesize Beam waist radius $\sigma$ [$\si{\micro\meter}$]};
\filldraw[color=white] (-1.9,-1.25) rectangle (-2.5,1.7);
\draw[] (-2.1,.25) node[rotate=90] {\footnotesize Relat. incoupled intensity};
\end{tikzpicture}
\caption{Modal intensity coupled into the fiber relative to the intensity of the incident beam for phase plate \protect \phasePl{1}.}
\label{fig:BeamSize1}
\end{figure}

As can be seen in figures \ref{fig:BeamSize0} and \ref{fig:BeamSize1}, it exists only one global maximum for the respective desired mode [$LP_{01}$ in figure \ref{fig:BeamSize0} and $LP_{11}$ in figure \ref{fig:BeamSize1}]. The same also holds for $LP_{02}$ using phase plate \phasePl{2} [see the appendix]. These maxima however do not necessarily coincide with the maxima of the total incoupled intensity. For the investigated few mode fiber [FMF] only if the phase shift $\Delta\Phi$ of the used phase plate was optimal [$\Delta\Phi=\SI{180}{\degree}$] and the beam waist radius was adapted respectively, then only the desired mode was excited.

As it can be seen in figure \ref{fig:BeamSize0}, using phase plate \phasePl{0} both $LP_{01}$ and $LP_{02}$ were excited. This shows that the used phase plates \phasePl{l} only discriminate between the mode groups $LP_{lm}$ and do not effect the latter index $m$. Thus if we would have used a fiber leading also higher modes from the mode groups $LP_{1m}$ and $LP_{2m}$, the incident power now lost at the coupling process would have excited these higher modes of the respective mode group.

In figure \ref{fig:BeamSize1} in addition to the optimal case of $\Delta\Phi=\SI{180}{\degree}$ also the case $\Delta\Phi=\SI{144}{\degree}$ is shown. The latter phase shift was derived from a first guess of the refractive index of the used material for the binary phase plates and the known difference in thickness $d$ [see figure \ref{fig:experimentalSetup}] and should be understood as a worst case consideration. Here one sees that the course of the graph for $LP_{0m}$ is exactly the same as in figure \ref{fig:BeamSize0}, though scaled corresponding to formula (\ref{eq:Two5}).

In table \ref{tab:BeamSize} the radii of the incident beam for optimal coupling into the respective mode together with the efficiency are summarized. The higher the index \phasePl{l} of the used phase plate, the larger the optimal beam waist radius becomes and the lower the efficiency. This is understandable considering on one hand the very high resemblance between a gaussian beam and a $LP_{01}$-mode and on the other hand that higher modes feature a larger mode diameter.

\begin{table}[H]
\caption{Beam waist radii of the incident beam for optimal coupling and the respective efficiency}
\label{tab:BeamSize}
\centering
\vspace*{6pt}
\begin{tabular}{ccc}
\parbox{5cm}{\centering phase plate} & \parbox{5cm}{\centering maximal coupling efficiency into the desired mode relative to the intensity of the incident beam} & \parbox{5cm}{\centering optimal beam waist radius for the incident beam}\\[15pt]
\phasePl{0} & $\SI{99.3}{\percent}$ & $\SI{10.0}{\micro\meter}$\\[3pt]
\phasePl{1} & $\SI{72.9}{\percent}$ & $\SI{12.7}{\micro\meter}$\\[3pt]
\phasePl{2} & $\SI{67.2}{\percent}$ & $\SI{15.1}{\micro\meter}$\\
\end{tabular}
\end{table}

Summarizing this, only for an optimal phase shift of the used phase plates an adjustment of the size of the incident beam by a simple intensity measurement after the fiber is possible for the investigated FMF. \\

\begin{figure}[H]
\centering
$\Delta\Phi = \SI{180}{\degree}$ \hspace*{6.25cm} $\Delta\Phi = \SI{144}{\degree}$\\[-6pt]
\begin{tikzpicture}[scale=1.1]
\draw (0,0) node {\includegraphics[width=.45\textwidth,trim = 0cm 0cm 3.5cm 0cm,clip]{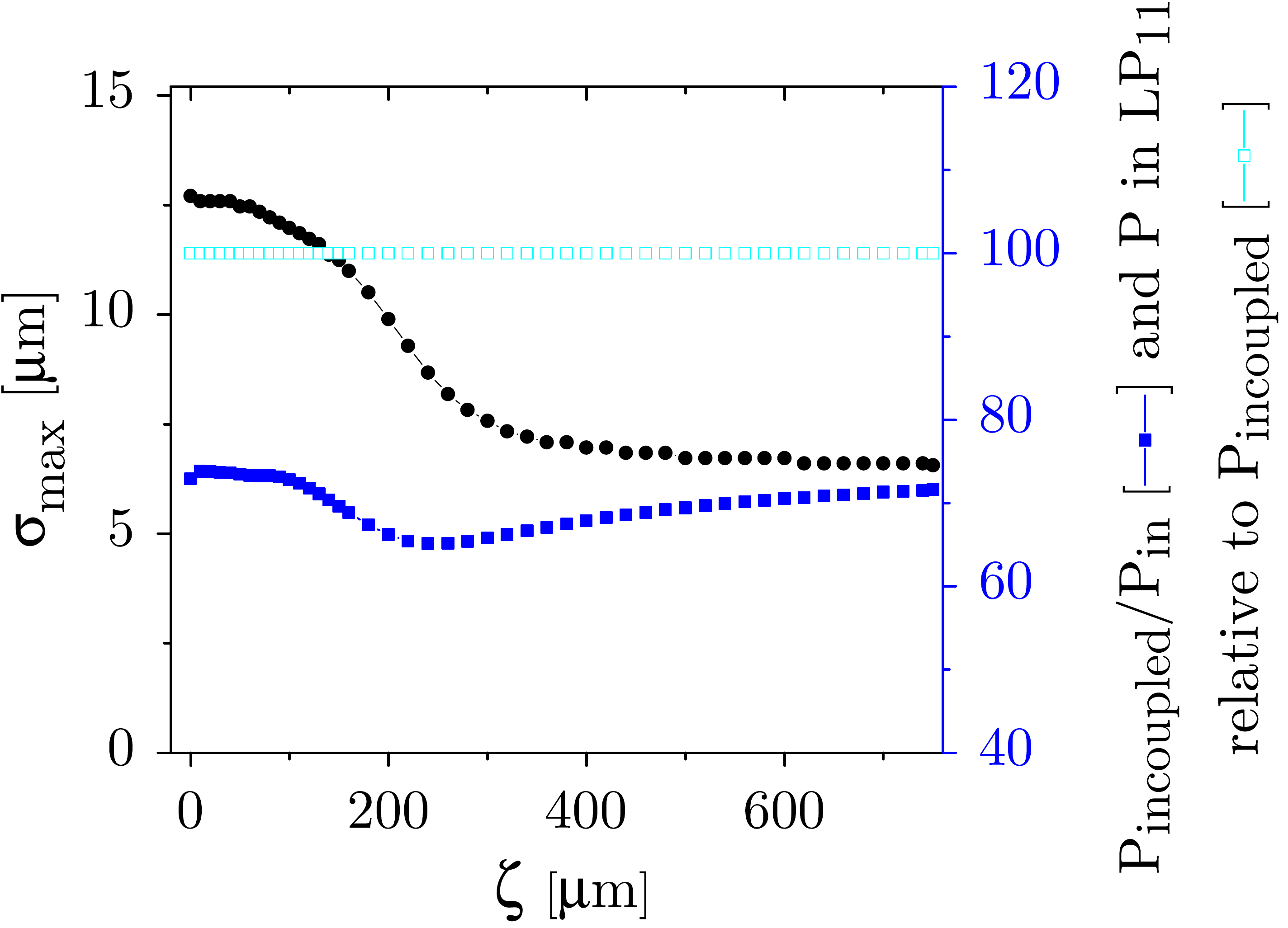}};
\filldraw[color=white] (3.8,-2.5) rectangle (3.2,2.8);
\draw[] (3.5,.25) node[rotate=90] {\footnotesize \textcolor{blue}{$\mathbf{P_{\text{ein}}/P_{\text{in}}}$} and \textcolor{cyan!80!white}{$P_{11}/P_{\text{ein}}$} [$\si{\percent}$]};
\end{tikzpicture}
\begin{tikzpicture}[scale=1.1]
\draw (0,0) node {\includegraphics[width=.45\textwidth,trim = 0cm 0cm 3.5cm 0cm,clip]{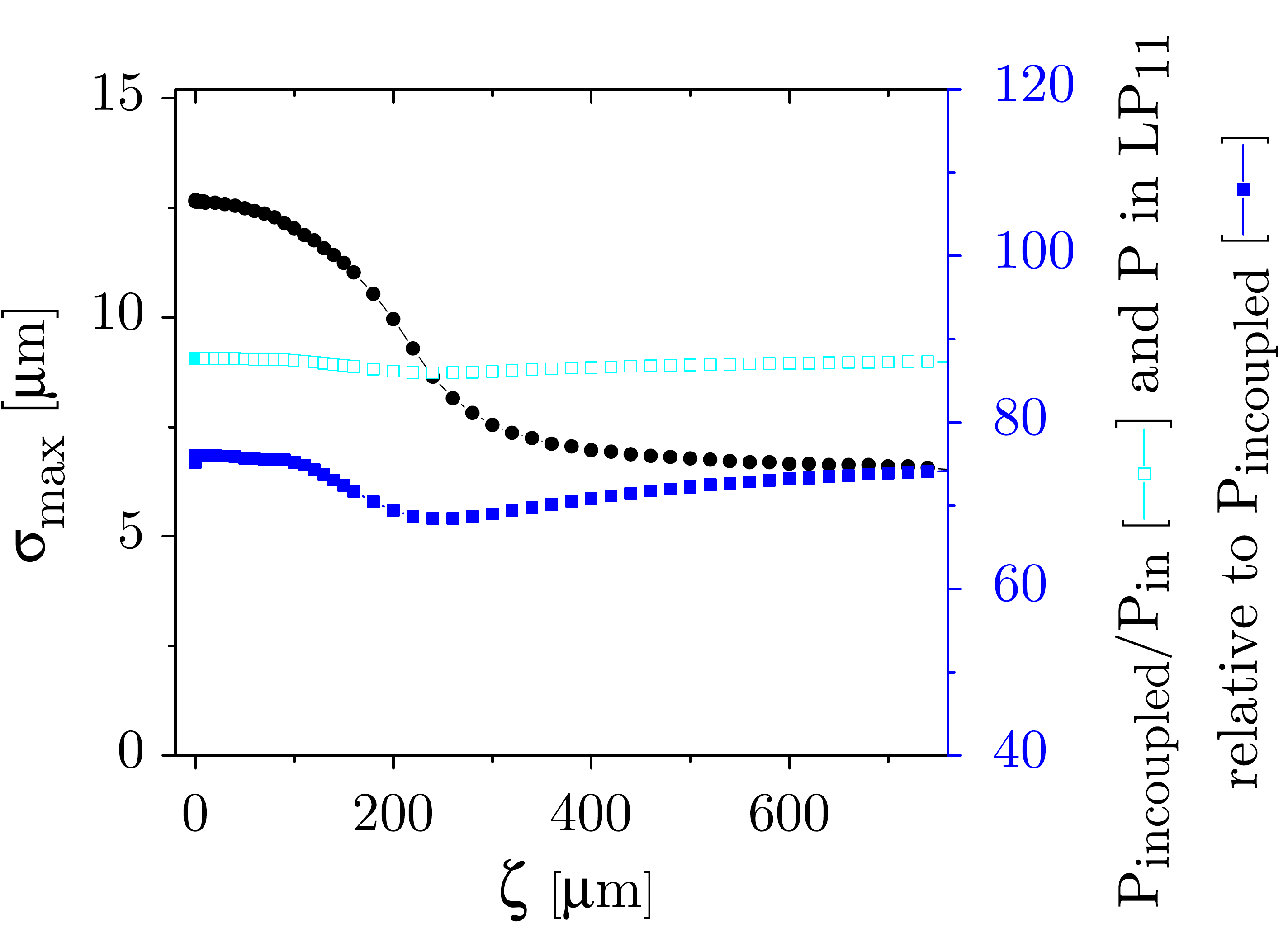}};
\filldraw[color=white] (3.8,-2.5) rectangle (3.2,2.8);
\draw[] (3.5,.25) node[rotate=90] {\footnotesize \textcolor{blue}{$\mathbf{P_{\text{ein}}/P_{\text{in}}}$} and \textcolor{cyan!80!white}{$P_{11}/P_{\text{ein}}$} [$\si{\percent}$]};
\end{tikzpicture}
\caption{Optimal beam waist radius $\sigma$ [left] and the intensity coupled into the fiber relative to the incident [\textcolor{blue}{${P_{\text{ein}}}/{P_{\text{in}}}$}] and the mode purity of $LP_{01}$ [\textcolor{cyan!80!white}{$P_{11}/P_{\text{ein}}$}] over the free space propagation distance $\zeta$ using phase plate \protect \phasePl{1}.}
\label{fig:FreeSpaceZeta1}
\end{figure}
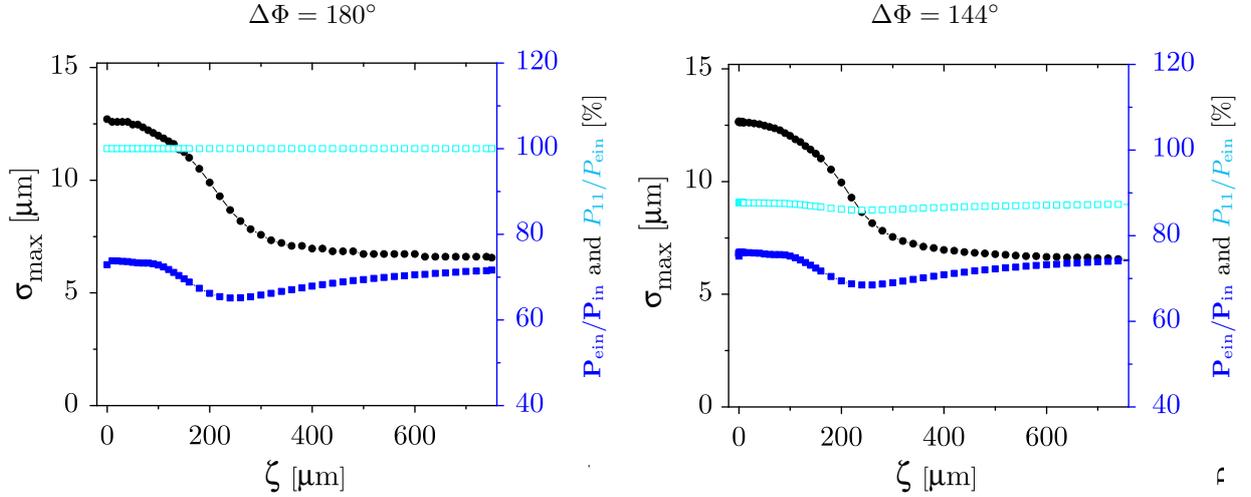

Furthermore, regarding the free space propagation distance $\zeta$ between the phase plate and the fiber input facet, for phase plate \phasePl{1} the dependency shown in figure \ref{fig:FreeSpaceZeta1} was calculated. The course of the graph for $LP_{02}$ using phase plate \phasePl{2} is quite similar and therefore only shown in the appendix. One should notice that the scale on the right side ranges from $\SI{40}{\percent}$ to $\SI{120}{\percent}$. The beam waist radius of the incident beam is adapted to each considered propagation distance $\zeta$. Thus a purity of the desired mode of approximately $\SI{100}{\percent}$ can be achieved numerically over the whole examined range for $\zeta$ from $\SI{0}{\micro\meter}$ to $\SI{750}{\micro\meter}$ for $\Delta\Phi=\SI{180}{\degree}$ [see $P_{11}/P_{\text{ein}}$]. For $\Delta\Phi=\SI{144}{\degree}$ the purity of $LP_{11}$ is a bit smaller than $\SI{90}{\percent}$ and varies slightly for different $\zeta$.

The larger the propagation distance, the more the beam diverges after the phase plate and thus has to be focused to a smaller spot initially. However it can be seen in figure \ref{fig:FreeSpaceZeta1}, that for a distance of less than $\zeta = \SI{100}{\micro\meter}$ the effect is quite small, independent of the phase shift $\Delta\Phi$. For the investigated setups $\zeta$ was assumed to be smaller than $\SI{50}{\micro\meter}$ and therefor these considerations disregarded.

\vspace*{12pt}
The last numerically considered property was the dependence of the modal coupling efficiency on the transversal displacement of the incoming beam relative to the concentric position of incident beam and fiber. The result for phase plate \phasePl{1} is shown in figure \ref{fig:TransDisp1}, for phase plate \phasePl{0} and phase plate \phasePl{2} in the appendix in figures \ref{fig:TransDisp0} and \ref{fig:TransDisp2}. Allthough the results may look similar to modal field distributions, they are, as said before, the modal intensities coupled into the fiber relative to the intensity of the incident beam.

As a quantitative measure for the sensitivity of this setup on the displacement of the incident fundamental Gaussian beam the full width at half maximum for the intensity coupled into the fiber of the respective desired mode was chosen; the values are shown in table \ref{tab:GaussInjectSens}.

\begin{table}[H]
\centering
\caption{Full width at half maximum for the respective desired mode in the monolithic setup.}
\vspace*{12pt}
\begin{tabular}{rccc}
									& \phasePl{0}	& \phasePl{1}	& \phasePl{2}\\[3pt]
$\Delta\Phi = \SI{144}{\degree}$	& \multirow{2}{*}{ $\SI{16.6}{\micro\meter}$ } 	& $\SI{19.2}{\micro\meter}$ 	& $\SI{24.6}{\micro\meter}$\\
$\Delta\Phi = \SI{180}{\degree}$	& 												& $\SI{18.2}{\micro\meter}$ 	& $\SI{24.6}{\micro\meter}$
\end{tabular}
\label{tab:GaussInjectSens}
\end{table} 

It is noteworthy that all of these values are smaller than the respective diameter of the incident Gaussian beam [\phasePl{0}: $2 \sigma_{\text{max}} = \SI{19.9}{\micro\meter}$, \phasePl{1}: $2 \sigma_{\text{max}} = \SI{25.4}{\micro\meter}$, \phasePl{2}: $2 \sigma_{\text{max}} = \SI{30.2}{\micro\meter}$]. To put this in perspective, it is mentioned that the fiber core diameter is $2 a = \SI{24.5}{\micro\meter}$.\\

\subsection{Experimental Results}

For the optical experiments the beam diameter of the incident Gaussian beam was adjusted to the optimal value for each experiment. The phase shift $\Delta\Phi$ and the free space distance between the phase plate and the fiber input facet could not be altered, but were determined by the setup. 

\begin{figure}[H]
\centering
\includegraphics[width=.95\textwidth, trim = 2cm 12cm 0.5cm 12.75cm, clip=true]{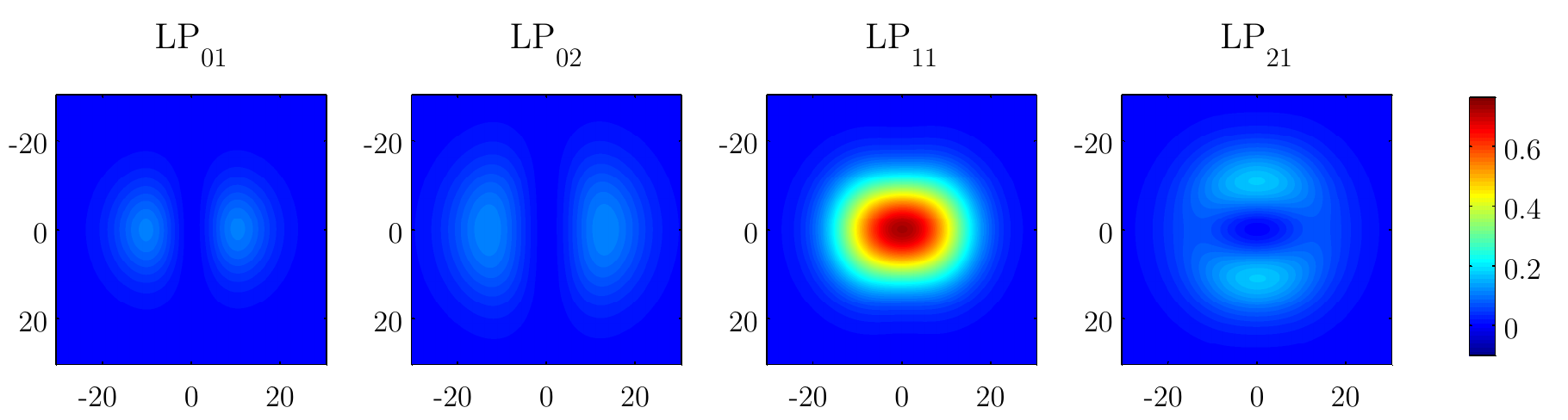}\\
{\hspace*{1cm} \includegraphics[width=.95\textwidth, trim = 2cm 12cm 0.5cm 12.75cm, clip=true]{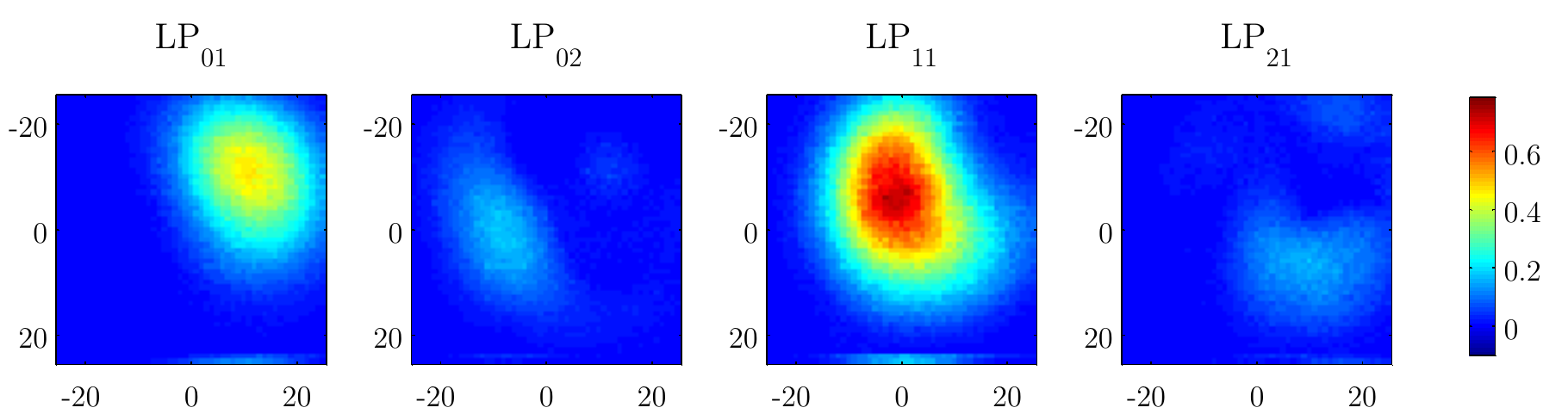}}
\caption{Modal coupling efficiency for the numerical [$\Delta\Phi=\SI{180}{\degree}$,top] and the optical [bottom] experiments regarding the transversal displacement of the incoming beam relative to the concentric position of incident beam and fiber for phase plate \protect \phasePl{1}.}
\label{fig:TransDisp1}
\end{figure}

Exemplarily the results are shown in figure \ref{fig:TransDisp1} for phase plate \phasePl{1}; the analogous results for phase plates \phasePl{0} and \phasePl{2} again can be found in the appendix [figures \ref{fig:TransDisp0} and \ref{fig:TransDisp2}].

The agreement of numerical and optical experiment is not perfect, but quite good. The main reason for this is the imperfect phase shift of the used binary phase plate. But also a possible tilt of the incident fundamental Gaussian beam could perturb the result.

\begin{figure}[H]
\centering
\scalebox{1}{\begin{tikzpicture}[x=.9cm,y=.02cm]
\draw (2.5,110) node[above] {\phasePl{0}};
\draw[thick,->] (.3,0) -- (4.7,0);  
\draw[thick,->] (.3,0) -- (.3,110);  

\foreach \y in {0,25,...,100} { 
	\draw[thick] (.38 ,\y) -- (.22,\y) node[anchor = east] {\small \y};
	}

\draw[] (-.75,50) node[rotate=90] {$\tilde{\rho}_{lm}^{\: 2}$ [$\si{\percent}$]};  

\foreach \mode/\x/\yTheo/\yExp in {
							$LP_{01}$		/1		/95.4		/93.3,
							$LP_{02}$		/2		/4.6		/3.5,
							$LP_{11}$		/3		/0		/3.1,
							$LP_{21}$		/4		/0		/.1
	}
	{
	\draw[fill=blue1] (\x,0) ++(.2,0) rectangle (\x-.2,\yExp);
	\draw[thick] (\x,4) -- (\x,0) node[anchor = north] {\small \mode};
	};
\end{tikzpicture}}
\hspace*{.25cm}
\scalebox{1}{\begin{tikzpicture}[x=.9cm,y=.02cm]
\draw (2.5,110) node[above] {\phasePl{1}};
\draw[thick,->] (.3,0) -- (4.7,0);  
\draw[thick,->] (.3,0) -- (.3,110);  

\foreach \y in {0,25,...,100} { 
	\draw[thick] (.38 ,\y) -- (.22,\y) node[anchor = east] {\small \y};
	}

		\draw[] (-.75,50) node[rotate=90] {$\tilde{\rho}_{lm}^{\: 2}$ [$\si{\percent}$]};  

	\foreach \mode/\x/\yTheo/\yExp in {
							$LP_{01}$		/1		/15.2		/8.7,
							$LP_{02}$		/2		/1.27		/7.7,
							$LP_{11}$		/3		/81.7		/80.2,
							$LP_{21}$		/4		/1.8		/3.4
	}
	{
	\draw[fill=blue1] (\x,0) ++(.2,0) rectangle (\x-.2,\yExp);
	\draw[thick] (\x,4) -- (\x,0) node[anchor = north] {\small \mode};
	};
\end{tikzpicture}}
\hspace*{.25cm}
\scalebox{1}{\begin{tikzpicture}[x=.9cm,y=.02cm]
\draw (2.5,110) node[above] {\phasePl{2}};
\draw[thick,->] (.3,0) -- (4.7,0);  
\draw[thick,->] (.3,0) -- (.3,110);  

\foreach \y in {0,25,...,100} { 
	\draw[thick] (.38 ,\y) -- (.22,\y) node[anchor = east] {\small \y};
	}

		\draw[] (-.75,50) node[rotate=90] {$\tilde{\rho}_{lm}^{\: 2}$ [$\si{\percent}$]};  

	\foreach \mode/\x/\yTheo/\yExp in {
							$LP_{01}$		/1		/14.5		/15.1,
							$LP_{02}$		/2		/.3		/0,
							$LP_{11}$		/3		/11.2		/11.6,
							$LP_{21}$		/4		/73.9		/73.3
	}
	{
	\draw[fill=blue1] (\x,0) ++(.2,0) rectangle (\x-.2,\yExp);
	\draw[thick] (\x,4) -- (\x,0) node[anchor = north] {\small \mode};
	};
\end{tikzpicture}}
\caption{Experimentally determined optimal values of the modal content coupled into the fiber for the investigated phase plates \protect \phasePl{0}, \protect \phasePl{1} and \protect \phasePl{2}.}
\label{fig:ExpOpt}
\end{figure}
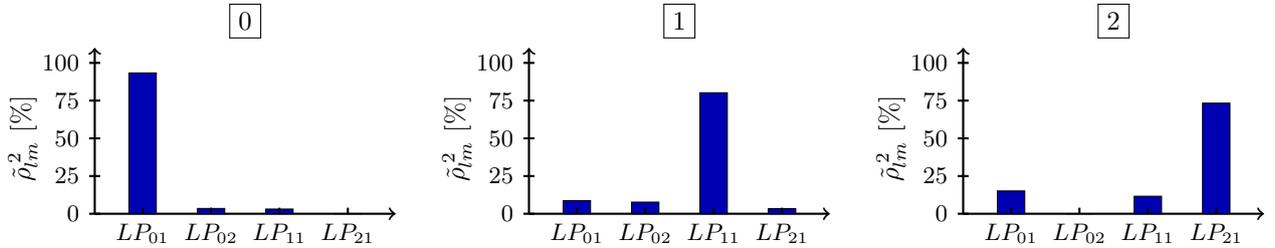

For all investigated phase plates the modal contents at the respective optimal positions are shown in figure \ref{fig:ExpOpt}. Here one sees the imperfect phase shift of the used phase plates even more clear. Nevertheless, the desired mode is in all three cases dominant.

\subsection{Comparison to the Free Space Setup}

All the previously mentioned experiments have been repeated using a \enquote{free space setup}, which is an allready well known approach [e.g. \cite{Hoy2013,Ste2011,Fla2013}]. Here the passive binary phase plate is not put directly in front of the fiber input facet but in the collimated laser beam between the laser and the first focusing lens $MO_1$ [see figure \ref{fig:experimentalSetup}].

This setup can also excite higher order modes with a purity of theoretically nearly $\SI{100}{\percent}$, provided that the phase shift $\Delta\Phi=\SI{180}{\degree}$. For the here used phase plates this was the case, as the phase plates were manufactured in a different manner 
using photolithography on a photo resist\footnote{AZ 1514H photoresist}, which's refractive index is well known, on top of a glass plate.

The free space setup showed to be more sensitive to a transversal displacement of the fiber relative to the concentric position of fiber and incident beam. This is reasonable, as a phase shaped beam features a faster variation of it's spatial profile than the unperturbed fundamental Gaussian beam.

\begin{figure}[H]
\centering
\parbox{7cm}{\centering \begin{tikzpicture}[x=1.4cm,y=.025cm]
\draw[thick,->] (.3,0) -- (3.7,0);  
\draw[thick,->] (.3,0) -- (.3,110);  

\foreach \y in {0,25,...,100} { 
	\draw[thick] (.38 ,\y) -- (.22,\y) node[anchor = east] {\y};
	}

\draw[] (-.55,50) node[rotate=90] {$\frac{P_{\text{incoupled}}}{P_{\text{in}}}$ [$\si{\percent}$]};  

\foreach \mode/\x/\yTheo/\yExpMon/\yExpFre in {
							\phasePl{0}		/1		/99		/50		/50,
							\phasePl{1}		/2		/73		/40		/25,
							\phasePl{2}		/3		/67		/30		/15
	}
	{
	\draw[fill=red] (\x,0) ++(-.375,0) rectangle (\x-.125,\yTheo);
	\draw[fill=blue1] (\x,0) ++(-.125,0) rectangle (\x+.125,\yExpMon);
	\draw[fill=green1] (\x,0) ++(.125,0) rectangle (\x+.375,\yExpFre);
	\draw[thick] (\x,4) -- (\x,0) node[anchor = north] {\mode};
	};
\end{tikzpicture}}
\hspace*{1cm} \parbox{7cm}{\centering \begin{tikzpicture}[x=1.4cm,y=.025cm]
\foreach \x/\y in {
	.6/60
	}
	{
	\draw (\x+.05,\y+45) rectangle (\x+3,\y-15);
	\filldraw[draw=black, fill=red] (\x+.25,\y+40) rectangle (\x+.5,\y+30);
	\draw (\x+.5,\y+35) node[anchor = west] {\small numeric};
	\filldraw[draw=black, fill=blue1] (\x+.25,\y+20) rectangle (\x+.5,\y+10);
	\draw (\x+.5,\y+15) node[anchor = west] {\small monolithic experiment};
	\filldraw[draw=black, fill=green1] (\x+.25,\y) rectangle (\x+.5,\y-10);
	\draw (\x+.5,\y-5) node[anchor = west] {\small free space experiment};
	};
\end{tikzpicture}}
\caption{Numerical and experimental [for the monolithic as well as the free space setup] optimal coupling efficiencies.}
\label{fig:DiagrammPrel}
\end{figure}
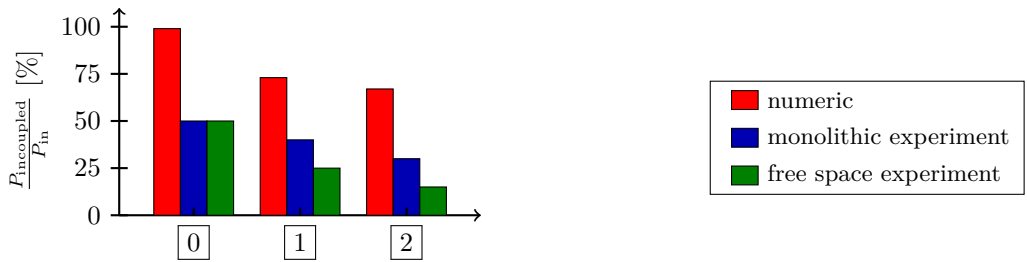

In figure \ref{fig:DiagrammPrel} the amount of the total into the fiber coupled relative to the incident power, as seen in the experiments, is depicted for the monolithic as well as the free space setup. These values were recorded at optimized coupling positions. Apparently the values for the monolithic setup are higher. One reason for this is that the phase shift $\Delta\Phi$ differed more from the optimal value of $\Delta\Phi=\SI{180}{\degree}$ for the monolithic setup and thus other modes could be excited, too; mainly $LP_{01}$. Yet the measured differences between the free space and the monolithic setup respectively are larger than explicable hereby and at least the experiments for phase plates \phasePl{0} and \phasePl{2} in the monolithic setup [see figures \ref{fig:ExpOpt}] showed, that the respective desired modes are very dominant. Therefore the main reason for the lower coupling efficiency of the non-monolithic setup is assumed to be the relatively long propagation distance between the free space phase plate and the fiber front facet, over which the phase shaped beam changes its profile.

\section{CONCLUSION}

We showed that the monolithic setup of fiber and phase plate [i.e. the phase plate being as near to the fiber input facet as possible] numerically can achieve a mode purity of $\SI{100}{\percent}$ in a few mode fiber. Experimentally the phase shift of the used phase plates was not optimal and thus the purity of the desired modes was smaller than $\SI{100}{\percent}$ [$\SI{93.3}{\percent}$ for $LP_{01}$, $\SI{80.2}{\percent}$ for $LP_{11}$ and $\SI{73.3}{\percent}$ for $LP_{21}$ with the respective phase plate]. In comparison with a setup, where the phase plate modified the collimated Gaussian beam before it was focused onto the fiber, it could be shown, that the monolithic approach was two times as efficient, regarding the incoupled power. Thus in order to use this setup in a real sensor application, at first the phase shift of the phase plate has to be optimized. Following this an investigation of the effects of pressure, stress or temperature on the modal content will be comfortably possible and thus a characterization as a possible sensor.

\appendix    

\section{Additional Graphs}

\vspace*{12pt}

\begin{figure}[H]
\centering
\hspace*{.75cm} $\Delta\Phi = \SI{180}{\degree}$ \hspace*{5.25cm} $\Delta\Phi = \SI{144}{\degree}$\\
\begin{tikzpicture}[scale=1.4]
\draw (0,0) node {\includegraphics[width=.35\textwidth, trim = .1cm 4.45cm 1.5cm 8.6cm, clip=true]{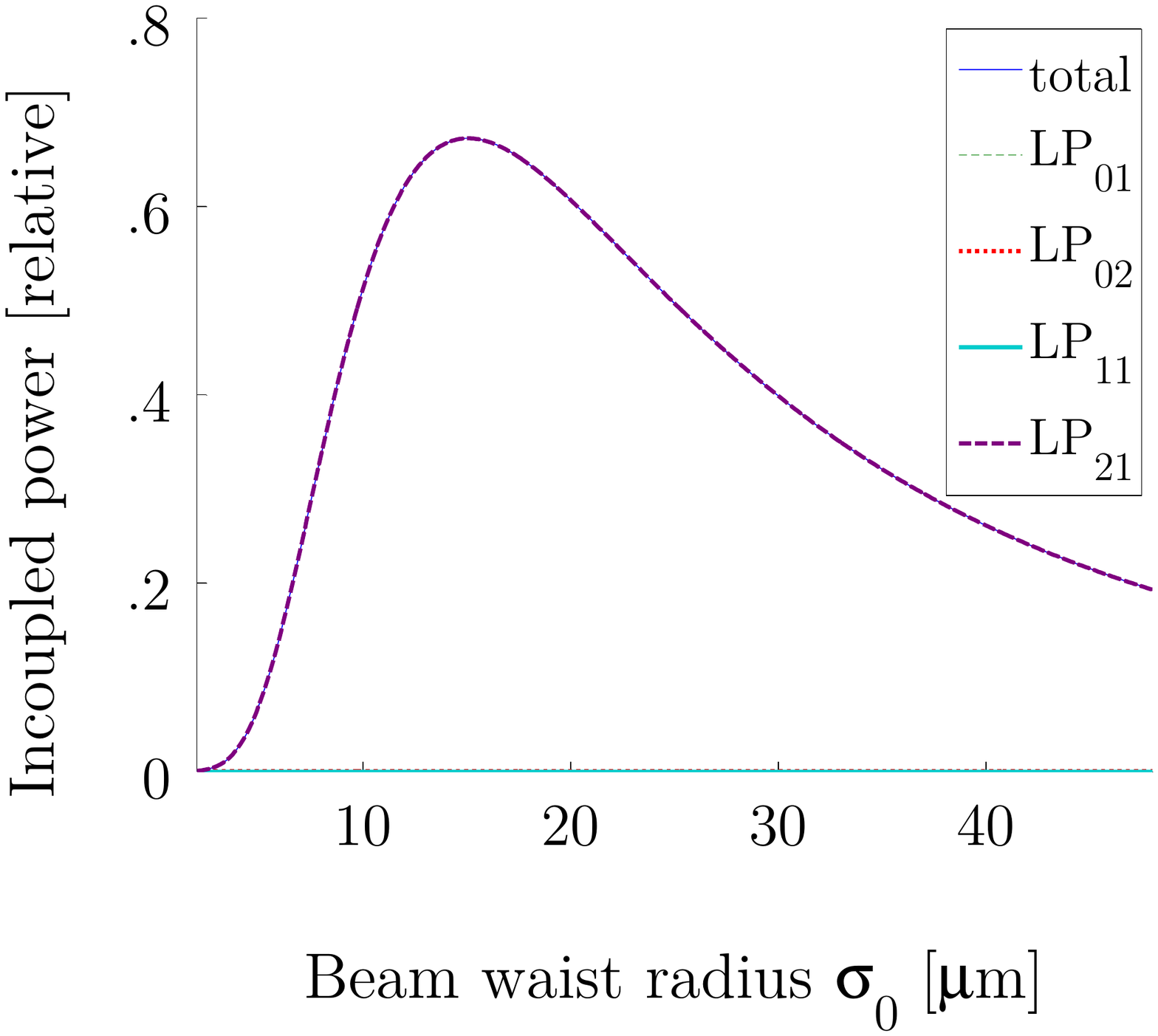}};
\filldraw[color=white] (-2,-1.5) rectangle (2,-2);
\draw (.35,-1.75) node {\footnotesize Beam waist radius $\sigma$ [$\si{\micro\meter}$]};
\filldraw[color=white] (-1.9,-1.25) rectangle (-2.5,1.7);
\draw[] (-2.1,.25) node[rotate=90] {\footnotesize Relat. incoupled intensity};
\end{tikzpicture} \hspace*{.5cm} \begin{tikzpicture}[scale=1.4]
\draw (0,0) node {\includegraphics[width=.35\textwidth, trim = .1cm 4.45cm 1.5cm 8.6cm, clip=true]{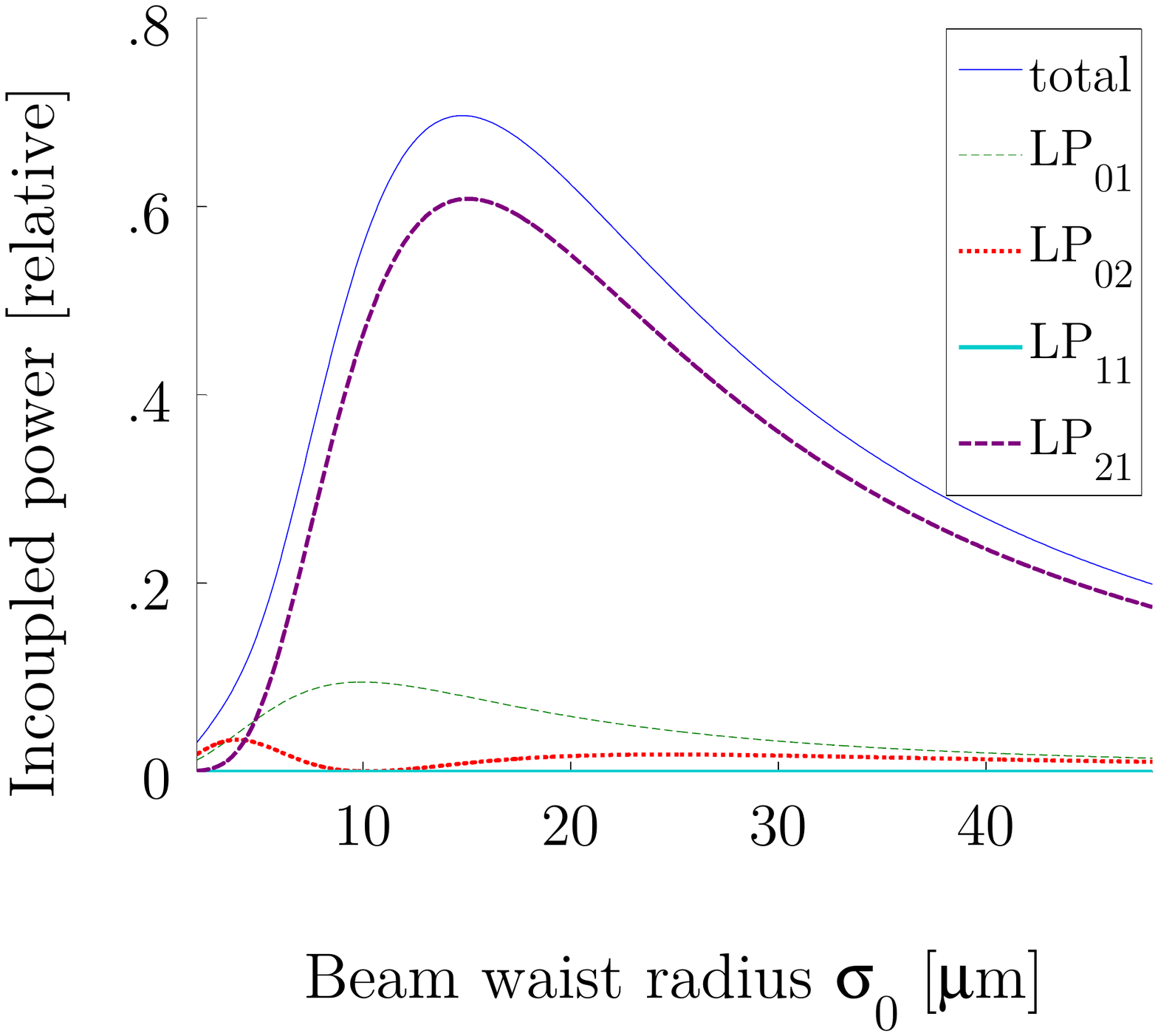}};
\filldraw[color=white] (-2,-1.5) rectangle (2,-2);
\draw (.35,-1.75) node {\footnotesize Beam waist radius $\sigma$ [$\si{\micro\meter}$]};
\filldraw[color=white] (-1.9,-1.25) rectangle (-2.5,1.7);
\draw[] (-2.1,.25) node[rotate=90] {\footnotesize Relat. incoupled intensity};
\end{tikzpicture}
\caption{Modal intensity coupled into the fiber relative to the intensity of the incident beam for phase plate \protect \phasePl{2}.}
\label{fig:BeamSize2}
\end{figure}

\vspace*{24pt}

\begin{figure}[H]
\centering
$\Delta\Phi = \SI{180}{\degree}$ \hspace*{6.25cm} $\Delta\Phi = \SI{144}{\degree}$\\[-6pt]
\begin{tikzpicture}[scale=1.1]
\draw (0,0) node {\includegraphics[width=.45\textwidth,trim = 0cm 0cm 3.5cm 0cm,clip]{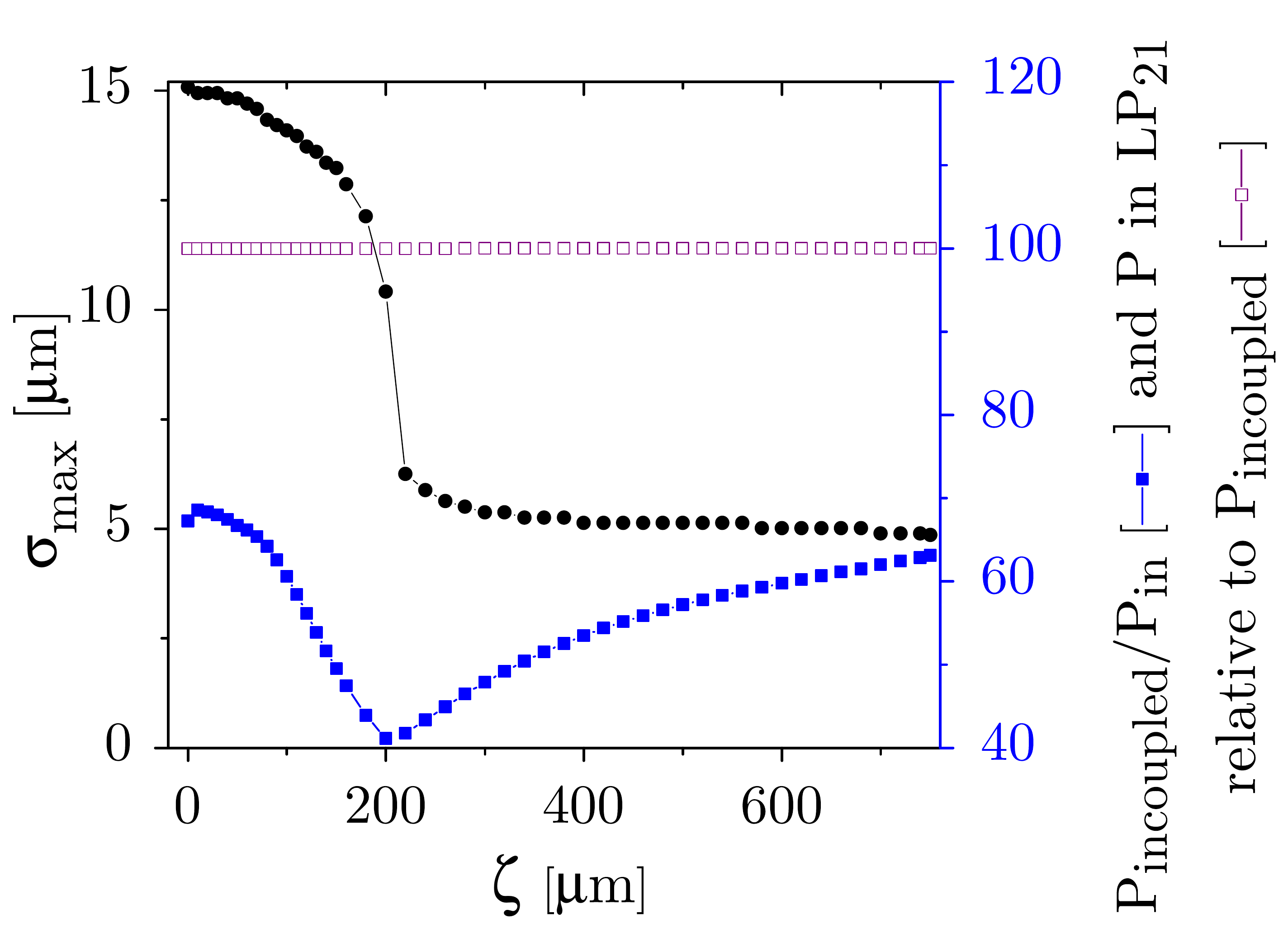}};
\filldraw[color=white] (3.8,-2.5) rectangle (3.2,2.8);
\draw[] (3.5,.25) node[rotate=90] {\footnotesize \textcolor{blue}{$\mathbf{P_{\text{ein}}/P_{\text{in}}}$} and \textcolor{violet!90!white}{$P_{21}/P_{\text{ein}}$} [$\si{\percent}$]};
\end{tikzpicture}
\begin{tikzpicture}[scale=1.1]
\draw (0,0) node {\includegraphics[width=.45\textwidth,trim = 0cm 0cm 3.5cm 0cm,clip]{./LMA-PzetaPhase1-matph144.pdf}};
\filldraw[color=white] (3.8,-2.5) rectangle (3.2,2.8);
\draw[] (3.5,.25) node[rotate=90] {\footnotesize \textcolor{blue}{$\mathbf{P_{\text{ein}}/P_{\text{in}}}$} and \textcolor{cyan!80!white}{$P_{11}/P_{\text{ein}}$} [$\si{\percent}$]};
\end{tikzpicture}
\caption{Optimal beam waist radius $\sigma$ [left] and the into the fiber coupled intensity relative to the incident [\textcolor{blue}{${P_{\text{ein}}}/{P_{\text{in}}}$}] and the mode purity of $LP_{02}$ [\textcolor{cyan!80!white}{$P_{21}/P_{\text{ein}}$}] over the free space propagation distance $\zeta$ using phase plate \protect \phasePl{2}.}
\label{fig:FreeSpaceZeta2}
\end{figure}

\begin{figure}[H]
\centering
\includegraphics[width=.95\textwidth, trim = 2cm 12cm 0.5cm 12.75cm, clip=true]{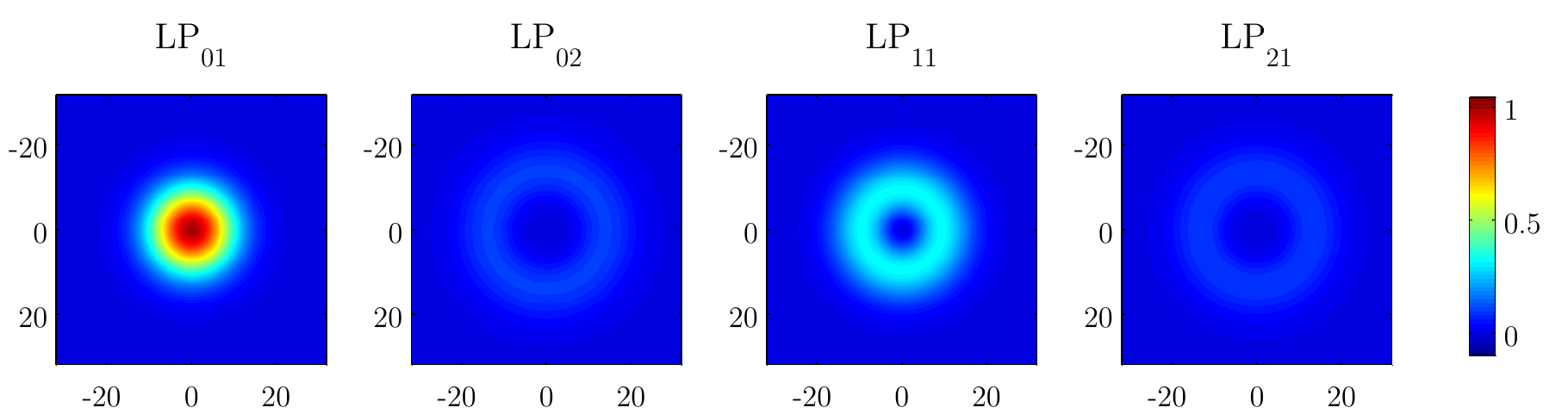}\\
{\hspace*{1cm} \includegraphics[width=.95\textwidth, trim = 2cm 12cm 0.5cm 12.75cm, clip=true]{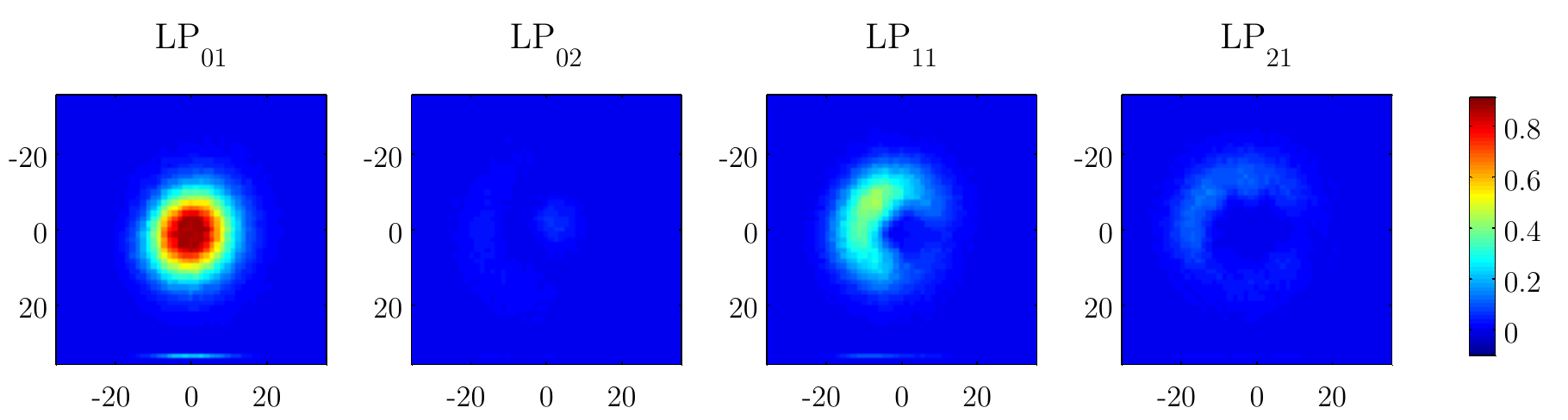}}
\caption{Modal coupling efficiency for the numerical [$\Delta\Phi=\SI{180}{\degree}$,top] and the optical [bottom] experiments regarding the transversal displacement of the incoming beam relative to the concentric position of incident beam and fiber for phase plate \protect \phasePl{0}.}
\label{fig:TransDisp0}
\end{figure}

\begin{figure}[H]
\centering
\includegraphics[width=.95\textwidth, trim = 2cm 12cm 0.5cm 12.75cm, clip=true]{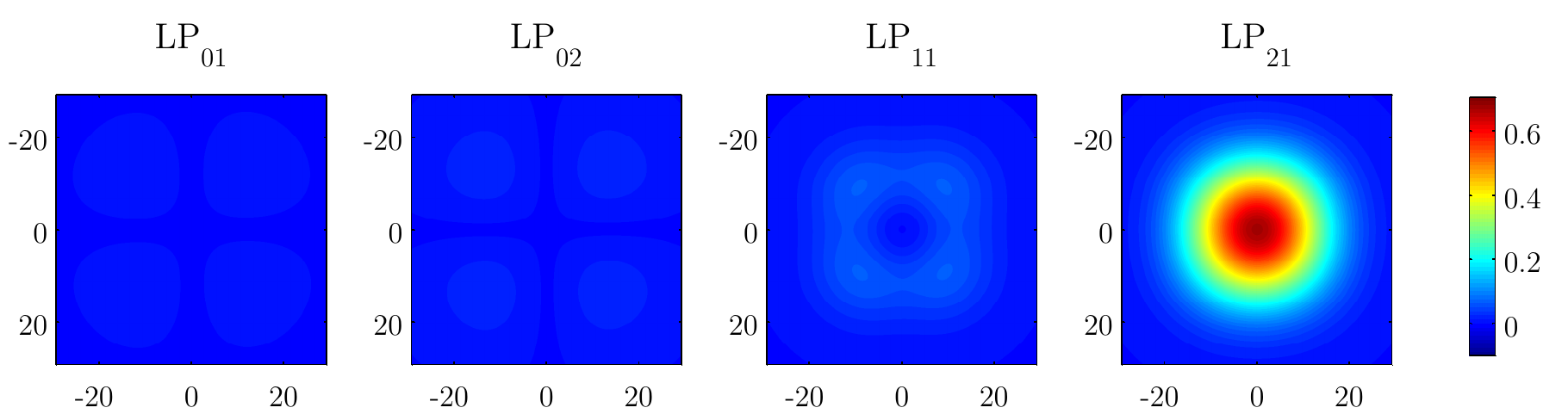}\\
{\hspace*{1cm} \includegraphics[width=.95\textwidth, trim = 2cm 12cm 0.5cm 12.75cm, clip=true]{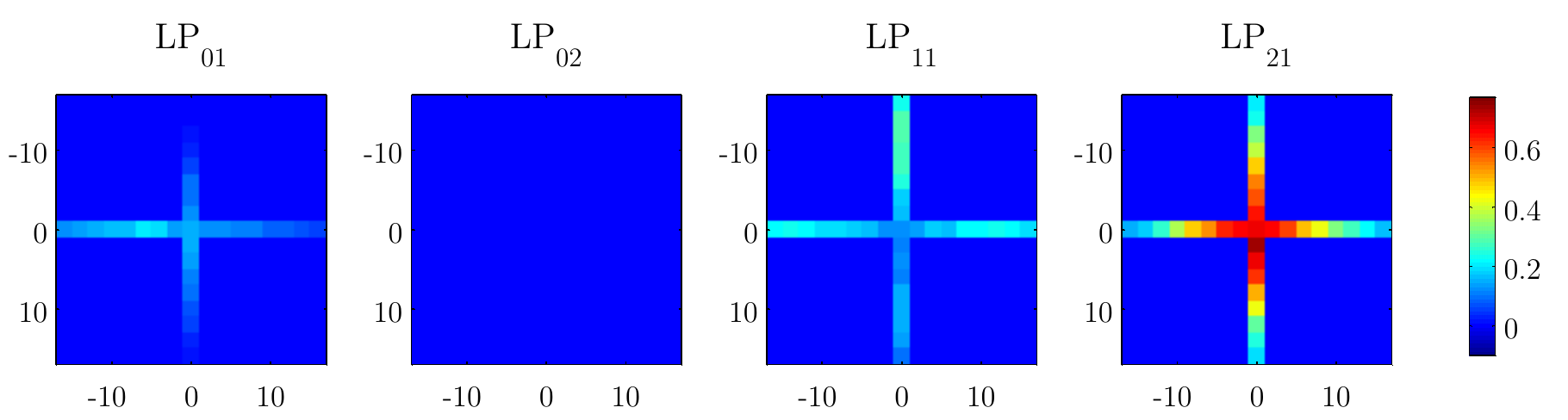}}
\caption{Modal coupling efficiency for the numerical [$\Delta\Phi=\SI{180}{\degree}$,top] and the optical [bottom] experiments regarding the transversal displacement of the incoming beam relative to the concentric position of incident beam and fiber for phase plate \protect \phasePl{2}}
\label{fig:TransDisp2}
\end{figure}

\acknowledgments     

The author would like to thank Mr. R. Pöhlmann from the IPHT Jena for his technical assistance. Without that this study would not have been possible.\\


\bibliography{library}   
\bibliographystyle{spiebibJo}   

\end{document}